%% file: spin_dmd.tex
\documentclass{siamltex}

\usepackage{amsfonts}
\usepackage{amssymb}
\usepackage{amsmath}
\usepackage{amsxtra}
\usepackage{latexsym}
\usepackage{verbatim}      
\usepackage{bm} 
\usepackage{graphicx}
\usepackage{float}
\usepackage{subfigure}
\usepackage{cite,color}
\usepackage[dvipsnames]{xcolor}
\usepackage{hyperref}

\input{macrosDMD.tex}

\numberwithin{equation}{section}

\title{Spin-Diffusions and Diffusive Molecular Dynamics\thanks{
This material is based upon work supported by the U.S. Department of
Energy Office of Science grant DE-SC0012733 and DE-SC0010549. Computational results reported here were obtained
on hardware supported by Drexel's University Research Computing
Facility, as well as the Minnesota Supercomputing Institute (MSI) at
the University of Minnesota.
}}

\author{%
Brittan A Farmer\thanks{%
School of Mathematics, University of Minnesota
Minneapolis, MN 55455, USA,
{\tt bafarmer@umn.edu}.}
\and
Mitchell Luskin\thanks{%
School of Mathematics, University of Minnesota
Minneapolis, MN 55455, USA,
{\tt luskin@math.umn.edu}.}
\and
Petr Plech\'a\v{c}\thanks{%
Department of Mathematical Sciences, University of Delaware,
Newark, DE 19716, USA,
{\tt plechac@math.udel.edu}.}
\and
Gideon Simpson\thanks{%
Department of Mathematics, Drexel University,
Philadelphia, PA USA,
{\tt simpson@math.drexel.edu}.}
}

\begin{document}
\maketitle
\begin{abstract}
  Metastable configurations in condensed matter typically fluctuate
  about local energy minima at the femtosecond time scale before
  transitioning between local minima after nanoseconds or
  microseconds.  This vast scale separation limits the applicability
  of classical molecular dynamics methods and has spurned the
  development of a host of approximate algorithms.  One recently
  proposed method is diffusive molecular dynamics which aims at
  integrating a system of ordinary differential equations describing
  the likelihood of occupancy by one of two species, in the case of a
  binary alloy, while quasistatically evolving the locations of the
  atoms. While diffusive molecular dynamics has shown to be efficient
  and provide agreement with observations, it is fundamentally a
  model, with unclear connections to classical molecular dynamics.

  In this work, we formulate a spin-diffusion stochastic process and
  show how it can be connected to diffusive molecular dynamics.  The
  spin-diffusion model couples a classical overdamped Langevin
  equation to a kinetic Monte Carlo model for exchange amongst the
  species of a binary alloy. Under suitable assumptions and
  approximations, spin-diffusion can be shown to lead to diffusive
  molecular dynamics type models.  The key assumptions and
  approximations include a well defined time scale separation, a
  choice of spin-exchange rates, a low temperature approximation, and
  a mean field type approximation.  We derive several models from
  different assumptions and show their relationship to diffusive
  molecular dynamics.  Differences and similarities amongst the models
  are explored in a simple test problem.\\

  Version: \today
\end{abstract}

\begin{keywords}
  Diffusive molecular dynamics, metastability, kinetic Monte Carlo,
  quasistationary distributions, mean field approximation
\end{keywords}

\section{Introduction}

Metals, alloys, and other condensed matter typically exhibit behavior
on at least two vastly different time scales making direct numerical
simulation by classical molecular dynamics (MD) impractical.  The
vibrational time scale of the atoms, measured in femtoseconds
($10^{-15}$~s), sets a first time scale, and constrains the time step
of MD.  Physically relevant phenomena occur on time scales of
microseconds ($10^{-6}$~s) or longer, setting a second time scale.
Milliseconds of the laboratory time can be achieved with direct MD
using special purpose hardware, but typical MD simulations only reach
nanoseconds~\cite{Voelz:2010hs}.

The source of this scale separation is the {\it metastability} present
in the physical system.  By metastability, we mean the tendency of the
system to persist in a particular arrangement, or conformation, of
atoms for a relatively long time before rapidly transitioning to some
other persistent conformation.  In many condensed matter problems,
these distinct metastable states correspond to local minimizers of a
potential energy landscape and transitions occur through saddle
points.  The system fluctuates about a local minimum on the fast time
scale, before a rare, but rapid, transition occurs.  As the feature of
physical interest is the sequence of transitions, classical MD will
exhaust computational resources resolving the fast, uninteresting,
fluctuations. Overcoming this scale separation has led to significant
investment in models and algorithms, including transition state theory
(TST), kinetic Monte Carlo (kMC), accelerated molecular dynamics, and
phase field crystal (PFC)~
\cite{perez2009accelerated,Voter:2002p12678,
  Xu:2008jk,Elder:2007kl,Wu:2010fc,Lelievre:2010uu,leimkuhlermd}.  All
of these methods exploit a ``fast/slow'' decomposition of the system,
averaging or approximating the fast femtosecond time scale, while
faithfully resolving details of the infrequent transitions between the
metastable regions.

{One method that has recently been formulated to overcome the
  time scale challenge is diffusive molecular dynamics (DMD)~
  \cite{Li:2011gn,Sarkar:2011wu,Sarkar:2012ce}.  While originally
  proposed based on physical grounds, here, we derive a DMD type
  system of equations through the following steps.  We begin with the
  overdamped Langevin equation, a diffusion, as a model for a binary alloy.  Some of the atoms
  in this system are species A and some are species B, and the type
  species asserts itself through the potential energy.  Metastability
  also bedevils overdamped Langevin, limiting the usefulness of 
  direct numerical integration. To circumvent the obstacle of metastability, we
  adjoin a spin-exchange kMC process, allowing for the two species, A
  and B, to swap.  This allows for more rapid evolution of the system
  than could be observed by direct numerical simulation of overdamped Langevin.
  We refer to this coupled process as the {\it spin-diffusion} model.}

Exploiting the time scale separation of the metastability using
idealized {\it quasistationary distributions} (QSDs), the fast
diffusional process averages out, leaving the kMC model with averaged
reaction rates.  This averaged kMC model then predicts the positions
of the atoms, as they are distributed by a distribution conditioned on
the current spin configuration.  Finally, an approximation of the
ensemble average of this scale separated kMC model gives rise to a
system of ordinary differential equations. These predict the
composition of the alloy and the structure of the distorted lattice.
It thus provides a tool for the study of mechanical deformation and
compositional evolution, as in the original DMD models.  Our
derivation, and the main goal of this work, bridges the gap between MD
and DMD, and constrains terms in DMD by relating them to our
underlying model.

The spin-diffusion model can be seen as a further generalization and
approximation of the weakly off-lattice kMC model of
\cite{Boateng:2014il}.  In that work, the authors quenched their
system to find the local minimizer, identified saddle points of the
associated basin, and then used Arrhenius reaction rates for the
associated kMC moves.  As we do not insist on quenching, our model
allows for finite temperature fluctuations about the local minima.
Additionally, while the method in \cite{Boateng:2014il} is stochastic,
requiring ensembles of simulations to study the evolution of averages,
a DMD type approach evolves an approximation of the time dependent
ensemble average.  This predicts the evolution of observables from a
single simulation.

DMD and spin-diffusion also bear resemblance to other models that have
appeared in the literature, including time dependent density
functional theory (TDDFT), phase field crystals (PFC), mean field
kinetic equations (MFKE), and compressible Ising~
\cite{Baker:1970cq,Gu:1996gi,Mitchell:2006ia,Marques:2004kw,Steinbach:2013gk,Gouyet:2003jw,Salinas:1973dl,Perez:2006ek,
  Fischer:1998ge}.  All of these models seek to predict scalar
quantities governing, for instance, probability of occupancy at a
point in space by a particular species. Time dependent quantities
evolve under systems of differential equations.

In contrast to TDDFT and PFC, DMD begins with the interatomic
potentials which would be used in a traditional MD simulation and
augments them to account for additional degrees of freedom.  Thus,
there is hope for better quantitative agreement.  DMD is entirely
classical in our formulation, and does not account for quantum
effects.  Also, unlike PFC, DMD explicitly retains atomistic detail
and is not a continuum limit.  Indeed, the DMD type models do not
correspond to taking a large-volume, fixed-density limit, as in many
mean field
approximations~\cite{Presutti:2009ha,huang2009introduction}.

Much, but not all, of the work on MFKE, whose ideas we draw from, has
been for on-lattice problems; DMD is explicitly off-lattice.  The
ideas which are closest to DMD include
\cite{Salinas:1973dl,Perez:2006ek} and related works, where the
lattice is permitted to fluctuate under harmonic approximations.  The
distinction with DMD, which also makes harmonic approximations, is
that its harmonic approximations are dynamically determined.

\subsection{Diffusive Molecular Dynamics}

The DMD method, as presented in the materials science literature,
begins by reformulating the MD problem in an extended state space.
Instead of having $N$ distinct atoms located at positions
$\bx = \{x_i\}_{i=1}^N$, one has $N$ distinct {\it atomic sites}
located at $\bx$ with associated scalar parameters
$\bs = \{s_i\}_{i=1}^N$.  The parameter $s_i\in [-1,1]$ captures the
likelihood of the site $i$ being occupied by the species A ($s_i = 1$)
or species B ($s_i = -1$).  In the earliest DMD works, elemental
materials were studied, and the scalar parameter was $c_i \in [0,1]$,
representing probability of occupancy by an atom as opposed to a
vacancy.  We emphasize the binary alloy problem and work in the $s_i$
coordinate.

A free energy, denoted by $\mathcal{F}$, is constructed as a function
of $\bs$ and the mean atomic positions $\bX=\{X_i\}_{i=1}^N$, along
with harmonic parameters, $\bk$, {defined in Section
  \ref{s:freeenergy}}.  First, $\mathcal{F}$ is minimized over $\bX$
and $\bk$ at fixed $\bs$:
\begin{equation} \label{e:Fmin1} (\bX, \bk) \in \argmin
  \mathcal{F}(\bs, \bX, \bk).
\end{equation}
This corresponds to a variational Gaussian (VG)
approximation~\cite{LeSar:1991ef}.  Hence, on the fast time scale, the
atomic sites, with a fixed composition, are vibrating about some well
defined configuration with the mean positions $\bX$.  On the slow time
scale, the composition $\bs$ evolves.  This minimization aspect of DMD
has recently been studied and posed in terms of relative entropy
(Kullback-Leibler divergence)~\cite{Simpson:2016aa}.  The variational
setting using the relative entropy was explored in \cite{KP2013} for
constructing a coarse-grained dynamics in a simpler model of
spin-diffusion on a fixed lattice describing non-Fickian diffusion
under external concentration gradient. However, DMD represents a
different approach in which the evolution of the coarse-grained model
is deterministic and is defined by a system of ordinary differential
equations.

The question is now how to evolve the composition in time.  In
\cite{Sarkar:2011wu}, one proposed model is driven by the free energy
gradients, {\begin{equation}\label{e:gradientflow1}
    \frac{ds_i}{dt} = \sum_{j\in \calN_i} m_{ij} \paren{\frac{\partial
        \mathcal{F}}{\partial s_j}-\frac{\partial
        \mathcal{F}}{\partial s_i} }.
  \end{equation}
  Here, $m_{ij}$ is a mobility tensor and $\calN_i$ is the set of
  atomic sites neighboring site $i$. Assuming the mobility is positive
  and symmetric ({\it i.e.}, $m_{ij} = m_{ji}$ and
  $\sum_{ij} v_i m_{ij} v_j>0$ for all vectors $v\neq 0$),} flow
\eqref{e:gradientflow1} conserves the total mass, $\sum s_i$, and has
$\mathcal{F}$ as a Lyapunov function.  Throughout these dynamics,
$(\bX, \bk)$ always instantaneously satisfy \eqref{e:Fmin1}, and thus
the evolution is {\it quasistatic}.

DMD has been used to study vacancy diffusion problems like sintering
of nanoparticles, diffusional void growth, and dislocation climb,
along with defect migration and binary alloys
\cite{Dontsova:2014et,Dontsova:2015fx,Li:2011gn,Sarkar:2012ce}.  More
generally, this method appears appropriate for modeling condensed
matter systems with an essential coupling between their composition
and their mechanical properties.

\subsection{Spin-Diffusion Dynamics}
\label{s:spinintro}

System \eqref{e:gradientflow1} is a modeling assumption, without a
direct relationship to an underlying MD model.  Furthermore, it
requires additional modeling of the mobility factor.  Here, we develop
a stochastic model which bridges MD to DMD, and relates coefficients,
like the mobility, to the underlying model.

For a binary alloy, each site has a position, $x_i$, and a ``spin'',
$\sigma_i\in \{\pm 1\}$, determining if it is species A
($\sigma_i=+1$) or B ($\sigma_i =-1$).  The ensemble averages of the
$\sigma_i$ will correspond to the $s_i$ of DMD.  Fixing the
composition at $\bsigma$, the positions evolve under the diffusion
\begin{equation}\label{e:diffusion1}
  d\bx(t) = - \nabla_x V(\bx(t), \bsigma)dt + \sqrt{2\beta^{-1}}d\bW(t),
\end{equation}
for a potential of both $\bx$ and $\bsigma$.  With a static value of
$\bsigma$, \eqref{e:diffusion1} corresponds to a standard MD method.

Fixing the configuration $\bx$, we now allow the composition to evolve
under a spin-exchange model with Kawasaki
dynamics~\cite{Kawasaki:1966kj,Kawasaki:1966wy,Krapivsky:2010ti}.
Recall that such a process is defined by reaction rates,
$r(\bsig\to \bsig';\bx)$, assumed to satisfy detailed balance with
respect to the Gibbs distribution, {\it i.e.},
\begin{equation}\label{e:rates1}
  r(\bsig\to \bsig'; \bx) e^{-\beta V(\bx, \bsig)} = r(\bsig'\to \bsig; \bx) e^{-\beta V(\bx, \bsig')}.
\end{equation}

Both of the above processes have $e^{-\beta V(\bx, \bsigma)}$ as the
invariant measure. If we let $L_x$ and $L_\sigma$ denote the
generators of the diffusion and spin-exchange processes, respectively,
then $L_\eps = \eps^{-1} L_x + L_\sigma$, induces a process that also
has $e^{-\beta V(\bx, \bsigma)}$ as the invariant measure. The
parameter $\eps$ captures the scale separation between the two
processes. In the $\eps\to 0$ limit, the positions, $\bx$, are
instantaneously distributed according to a distribution parametrized
by the composition, $\bsigma$, while the composition evolves under the
spin dynamics with rates averaged over $\bx$.  After then
approximating the ensemble average, we obtain a system of equations
with many of the same features as DMD.  {The actual coupling of
  the processes is more subtle, but this is the overarching idea.}

\subsection{Outline}

This work is organized as follows.  In Section~\ref{s:state}, we
present the state space that underlies these models and the associated
potentials.  The essential ideas of DMD are then reviewed in
Section~\ref{s:dmd}.  After that, we present our stochastic
spin-diffusion model in Section \ref{s:spindiffusion}.  Numerical
examples are given in Section~\ref{s:numerics} and we conclude with a
discussion in Section~\ref{s:discussion}.  Additional calculations and
details appear in the Appendix.

\section{State Space and Potential Energies}
\label{s:state}

The framework of both DMD and the spin-diffusion model begins with the
state space $(\bx, \bsigma) \in \Omega^{ N}\times \{\pm 1\}^N$.  The
domain $\Omega \subset \R^d$ may be all of $\R^d$ or a subset with
boundary conditions.  The pair $(x_i, \sigma_i)$ identifies the
position and species of the $i$-th site.  Pair potential interactions
can then be defined as
\begin{equation}
  \label{e:pair}
  V(\bx,\bsigma) = \sum_{i<j} \phi_{\sigma_i, \sigma_j}(|x_i - x_j|).
\end{equation}
Pair interaction between two atomic sites thus depend on both the
species at the sites and the distance between them.  The reader can
imagine more general potentials of this form, such as EAM.  A key
assumption that we make on the potential is
\begin{assumption}[Permutation Invariance of Potential]
  \label{asm:invariance}
  Given $(\bx,\bsigma)$, if we permute the indices of both $\bx$ and
  $\bsigma$ in the same way to obtain $(\bx', \bsigma')$, then
  $V(\bx, \bsigma) = V(\bx', \bsigma')$.
\end{assumption}\\
Given such a potential, we define the distribution
\begin{equation}
  \label{e:mudist1}
  \mu(d\bx, \bsigma) = Z^{-1} \exp(-\beta V(\bx, \bsigma))d\bx, \quad
  Z = \sum_{\bsigma}{}^{'}\int_{\Omega^N}  \exp(-\beta V(\bx, \bsigma))d\bx.
\end{equation}
The summation is primed to indicate that the net composition is
constrained:
\begin{equation}
  \label{e:mass}
  \sum_{i=1}^N \sigma_i = M, \quad N_{\rm A} + N_{\rm B} = N, \quad
  N_{\rm A} -
  N_{\rm B} = M.
\end{equation}
$N_{\rm A}$ and $N_{\rm B}$ are fixed beforehand.  The partition
function $Z$ suffers from both high dimensionality and a combinatorial
explosion.  Notice that because of the invariance in Assumption
\ref{asm:invariance}, for any fixed $\bsigma_\star$ satisfying
\eqref{e:mass},
\begin{equation}
  \label{e:Zrelation1}
  Z = \sum_{\bsigma}{}^{'}\int_{\Omega^N}  \exp(-\beta V(\bx,
  \bsigma))d\bx = {{N}\choose{N_{\rm A}}} \int_{\Omega^N}  \exp(-\beta V(\bx,
  \bsigma_\star))d\bx.
\end{equation}
{Thus, any process which samples $e^{-\beta V(\bx, \bsigma)}$
  over $\R^{d\cdot N} \times \{\pm 1\}^N$ can be used to compute
  ensemble averages over $e^{-\beta V(\bx, \bsigma_\star)}$ of
  observables which are also invariant to permutations, such as the
  energy.  We also assume:
  \begin{assumption}[Finitely Many Minima of Potential]
    \label{asm:finiten}
    For a given $\bsigma$, $V(\bx, \bsigma)$ has a finite number of
    minima in $\Omega^N$.
  \end{assumption}\\
  By Assumption \ref{asm:invariance}, the number of minima is then
  invariant to $\bsigma$.  The minima have corresponding basins of
  attraction, $D_\ell(\bsigma)$.  Excluding the measure zero set
  corresponding to points which converge to saddles, which we will not
  give further consideration to, these open sets disjointly partition
  the state space, up to the measure zero set.  The union of their
  closures, $\bar D_\ell(\bsigma)$, cover $\bar\Omega^{N}$; for each
  $\bsigma$ satisfying \eqref{e:mass},
  \begin{equation}
    \label{e:domaindecomp}
    \bar\Omega^{N} = \cup_{\ell=1}^{n} \bar D_\ell(\bsigma), \quad
    D_\ell(\bsigma)\cap D_{\ell'}(\bsigma) = \emptyset \quad \text{if
      $\ell\neq \ell'$}.
  \end{equation}
  The partition function, from \eqref{e:Zrelation1} can be decomposed
  as
  \begin{equation}
    \label{e:Zrelation2}
    Z = \sum_{\bsigma}{}^{'}\sum_{\ell=1}^n \int_{D_\ell(\bsigma)} \exp(-\beta
    V(\bx, \bsigma))d\bx = \sum_{\bsigma}{}^{'}\sum_{\ell=1}^n
    Z_\ell(\bsigma).
  \end{equation}
  The $Z_\ell(\bsigma)$ are the partition functions of
  $1_{D_\ell(\bsigma)}(\bx)e^{-\beta V(\bx, \bsigma)}$, the {\it
    restricted} Gibbs distributions.}

{The potential we examine in this work} includes both pair
potential interactions and self-interaction trapping terms:
\begin{equation}
  \label{e:modelV1}
  \begin{split}
    V(\bx, \bsigma) &= \sum_{i<j} \phi_{\sigma_i, \sigma_j}(|x_{ij}|)
    +
    \sum_{i} u_{\sigma_i}(|x_i|)\\
    & = \sum_{i<j}
    \tfrac{1}{4}(1+\sigma_i)(1+\sigma_j)\phi_{++}(|x_{ij}|)\\
    &\quad
    +\tfrac{1}{4}[(1+\sigma_i)(1-\sigma_j)+(1-\sigma_i)(1+\sigma_j)]\phi_{+-}(|x_{ij}|)\\
    & \quad + \tfrac{1}{4}(1-\sigma_i)(1-\sigma_j)\phi_{--}(|x_{ij}|)\\
    & \quad + \sum_i
    \tfrac{1}{2}(1+\sigma_i)u_+(|x_i|)+\tfrac{1}{2}(1-\sigma_i)u_-(|x_i|),
  \end{split}
\end{equation}
where $x_{ij} \equiv x_i-x_j$. Expression~\eqref{e:modelV1} can be
recast as
\begin{equation}
  \label{e:modelV2}
  \begin{split}
    V(\bx, \bsigma)& = \sum_{i<j} -\sigma_i \sigma_j J_{ij}(\bx) +
    \sum_i \sigma_i h_i(\bx) +f(\bx).
  \end{split}
\end{equation}
See Appendix \ref{a:modelV} for details of $J_{ij}$, $h_i$ and $f$.

As noted in the introduction, the stochastic process we develop
involves exchanges of spins between atomic sites.  Given composition
$\bsigma$, a spin-exchange will be denoted $\bsigma^{ij}$, with
components
\begin{equation}
  \label{e:exchange1}
  \sigma^{ij}_k  = \begin{cases}
    \sigma_i & k = j,\\
    \sigma_j & k = i,\\
    \sigma_k & k\neq i,j.
  \end{cases}
\end{equation}
Viable exchanges will be limited to a set, $\calN_i$, of neighboring
sites, which may change in time.  With fixed configuration $\bx$, the
change in the potential energy, which drives the exchanges, is
\begin{equation}
  \label{e:deltaV}
  \begin{split}
    \Delta_{ij} V(\bx, \bsigma) &\equiv V(\bx, \bsigma^{ij}) - V(\bx,
    \bsigma)\\
    &\quad = (\sigma_j - \sigma_i) \Big\{-\sum_{k\neq i,j}
    (J_{ik}(\bx) - J_{jk}(\bx) )\sigma_k + h_i(\bx) - h_j(\bx)\Big\}.
  \end{split}
\end{equation}

\section{DMD Model}\label{s:dmd}

Here we briefly summarize the ingredients of DMD.  We refer the reader
to \cite{Simpson:2016aa} for a more detailed treatment.

\subsection{Free Energy Approximation}
\label{s:freeenergy}

First, an approximate free energy is developed.  Towards this,
constraint \eqref{e:mass} is relaxed via a change of ensemble:
\begin{assumption}[Change of Ensemble]
  \label{asm:meanmass}
  Mass constraint \eqref{e:mass} only holds on average; the mean spin
  values satisfy the total mass constraint,
  \begin{equation}
    \label{e:constraints}
    \E[\sigma_i] = s_i, \quad \sum_{i=1}^N s_i = M.
  \end{equation}
  The mean values, $s_i$, are enforced by the Lagrange multipliers,
  $\lambda_i$, for the distribution
  \begin{equation}
    \label{e:nudist1}
    \begin{split}
      \nu(d\bx, \bsigma) = Z_\nu^{-1} \exp(-\beta V(\bx, \bsigma)+\beta\blambda \cdot\bsigma)d\bx, \\
      Z_\nu = \sum_{\bsigma} \int_{\Omega^N} \exp(-\beta V(\bx,
      \bsigma)+\beta\blambda \cdot\bsigma)d\bx.
    \end{split}
  \end{equation}
\end{assumption}

The summation in \eqref{e:nudist1} is no longer primed as the
$\bsigma$ varies over all of $ \{\pm 1\}^{N}$.  To proceed further the
aforementioned VG approximation is made~\cite{LeSar:1991ef}.  First,
we define $\tilde \nu$ as
\begin{equation}
  \label{e:nudist2}
  \tilde \nu(d\bx, \bsigma;\bX, \bk)  =\tilde Z_\nu^{-1} \exp(-\beta \tilde V(\bx;\bX,\bk)
  + \beta \tilde \blambda\cdot \bsigma)d\bx,
\end{equation}
with harmonic potential
\begin{equation}
  \label{e:harmonic1}
  \tilde V(\bx;\bX,\bk) =  \sum_{i=1}^N \frac{k_i}{2}|x_i - X_i|^2.
\end{equation}
{Note, a variable with a $\tilde{}$ on it
  indicates it is associated with a VG approximation.}   For potential
\eqref{e:harmonic1}, many of the terms in~\eqref{e:nudist2} become explicit:
\begin{align}
  \label{e:ZVG1}
  \tilde Z_\nu &= \prod_{i=1}^N 2\tilde Z_i \cosh(\beta \tilde\lambda_i)=
                 \prod_{i=1}^N \frac{2\tilde Z_i}{\sqrt{1-s_i^2}},\\
  \label{e:ZVG2}
  \tilde Z_i &= \int_\Omega \exp\set{-\frac{\beta k_i}{2}|x_i-X_i|^2}dx_i,\\
  \label{e:lagrangeVG1}
  \tilde\lambda_i & =\beta^{-1} \arctanh (s_i).
\end{align}
$\bX$ and $\bk$ are then chosen to minimize the relative entropy
between $\tilde\nu$ and $\nu$~\cite{Simpson:2016aa}.  {Recall
  that the relative entropy metric between two probability measures,
  $\mu$ and $\nu$, on a common state space is given by
  \begin{equation}
    \label{e:R}
    \mathcal{R}(\mu|| \nu) = \begin{cases}
      \E^{\mu}\bracket{\log\frac{d\mu}{d\nu}} = \int
      \log\frac{d\mu}{d\nu}d\mu, & \mu\ll \nu, \\
      \infty, & \text{otherwise}.
    \end{cases}
  \end{equation}
  Here, $\mu\ll \nu$ indicates that $\mu$ is {\it absolutely
    continuous} with respect to $\nu$.  $\mathcal{R}$ is non-negative
  and vanishes if and only if $\mu = \nu$.  For our problem,
  \begin{equation}
    \label{e:relent0}
    \mathcal{R}(\tilde \nu|| \nu) = \beta \E^{\tilde \nu}[\Delta V] -
    \beta \Delta \blambda \cdot \mathbf{s} + \log Z_\nu - \log \tilde
    Z_\nu, \quad \Delta V = V - \tilde V, \quad \Delta \blambda = \blambda - \tilde \blambda.
  \end{equation}
  We thus make the approximation:}
\begin{approximation}[Variational Gaussian/Relative Entropy
  Approximation]
  \label{apx:VG}
  \newline Replace $\nu$ by $\tilde \nu$ in the model by solving
  {\begin{equation}
      \label{e:relent1}
      (\bX, \bk)\in  \argmin\mathcal{R}(\tilde \nu|| \nu)
    \end{equation}}
  and then employ the free energy upper bound $\mathcal{F}$ in the
  dynamics:
  \begin{equation}
    \label{e:freebound}
    -\beta^{-1} \log Z_\nu + \blambda \cdot \bs \leq  \E^{\tilde
      \nu}[\Delta V] -\beta^{-1} \log \tilde Z_\nu + \tilde\blambda \cdot \bs =\mathcal{F}.
  \end{equation}

\end{approximation}

We simplify the energy expression, $\mathcal{F}$, for later
comparison.  First, let
\begin{equation}
  \label{e:muconfig1}
  \tilde \mu(d\bx;\bX,\bk) = \prod_{i=1}^N
  {\tilde Z_i^{-1}\exp\set{-\frac{\beta
        k_i}{2}|x_i-X_i|^2}dx_i}=
  \prod_{i=1}^N \tilde\mu_i(dx_i; X_i, k_i),
\end{equation}
corresponding to the configurational portion of the distribution.
Next, define
\begin{equation}
  \label{e:dmdavgterms}
  \tilde J_{ij}(\bX,\bk) = \E^{\tilde\mu}[J_{ij}(\bx)], \quad \tilde
  h_{i}(\bX,\bk) = \E^{\tilde\mu}[h_{i}(\bx)],\quad \tilde f(\bX, \bk) = \E^{\tilde\mu}[f(\bx)],
\end{equation}
where $J$, $h$, and $f$ are as in \eqref{e:modelV2}.  With these
definitions,
\begin{equation}
  \label{e:F1}
  \begin{split}
    \mathcal{F} &= \sum_{i<j} -s_i s_j \tilde J_{ij}(\bX, \bk) +
    \sum_i s_i \tilde h_i(\bX, \bk) + \tilde f(\bX, \bk) - \sum_{i}
    \frac{k_i}{2}\E^{\tilde \mu_i}[|x_i - X_i|^2]
    \\
    &\quad +\beta^{-1} \sum_i \frac{1+s_i}{2}\ln\frac{1+s_i}{2} +
    \frac{1-s_i}{2}\ln\frac{1-s_i}{2} - \ln \tilde Z_i.
  \end{split}
\end{equation}
Thus,
\begin{equation}
  \label{e:Fgrad}
  \frac{\partial \mathcal{F}}{\partial s_i} = - \sum_{j} s_j\tilde J_{ij}(\bX,
  \bk) + \tilde h_i(\bX, \bk) + \beta^{-1}\arctanh(s_i).
\end{equation}

\subsection{Evolution Equations}
\label{s:dmdflow}

To evolve $\bs$ in time under \eqref{e:gradientflow1}, modeling
assumptions are needed for the mobility.  Simple mobility models
include
\begin{align}
  \label{e:constmobility}
  \text{Constant mobility: } &  m_{ij} =\begin{cases} m_{\rm c} & j \in N_i,\\
    0 & j \notin N_i.
  \end{cases}\\
  \label{e:sarkarmobility}
  \text{Rate limited mobility: } & m_{ij} = \begin{cases}
    m_{\rm r} (1+s_i) (1-s_j), & j \in N_i, \quad \frac{\partial\mathcal{F}}{\partial{s_i}}>
    \frac{\partial\mathcal{F}}{\partial{s_j}},\\
    m_{\rm r} (1-s_i) (1+s_j), & j \in N_i, \quad \frac{\partial\mathcal{F}}{\partial{s_i}}<
    \frac{\partial\mathcal{F}}{\partial{s_j}},\\
    0 & j \notin N_i.
  \end{cases}
\end{align}
The constant mobility model is simplest, but the rate limited
mobility, from \cite{Sarkar:2011wu}, accounts for limitations to mass
transfer due to the amount of A and B at each site.

An alternative to \eqref{e:gradientflow1} is the so-called ``Master
Equation,'' which appeared in
\cite{Li:2011gn,Sarkar:2011wu,Sarkar:2012ce,
  Dontsova:2014et,Dontsova:2015fx},
\begin{equation}
  \label{e:master1}
  \frac{ds_i}{dt} = \sum_{j\in N_i} (1+s_j)(1-s_i)\Gamma_{j\to i} - (1+s_i)(1-s_j)\Gamma_{i\to j}.
\end{equation}
The $\Gamma_{j\to i}$ are the rates of mass transfer from the site $j$
to site $i$.  They are explicitly modulated by steric factors,
ensuring that there is both mass available to transfer from $j$ and
space available at $i$.  In \cite{Sarkar:2011wu}, \eqref{e:master1}
was preferred to \eqref{e:gradientflow1} due to reduced numerical
stiffness, and it has been used in
\cite{Li:2011gn,Sarkar:2011wu,Sarkar:2012ce,
  Dontsova:2014et,Dontsova:2015fx}.

A common choice for the jump rate is
\begin{align}
  \label{e:jumprates1}
  \Gamma_{j\to i} &=\begin{cases}
    \kappa \exp\big\{- \beta \Qm- \beta (f_i - f_j)\big\}
    & j \in N_i,\\
    0 & j \notin N_i,
  \end{cases}
  \\
  \label{e:formationenergy1}
  f_i &= \frac{\partial\mathcal{F}}{\partial{s_i}} -
        \beta^{-1}\arctanh(s_i)= - \sum_{j} s_j\tilde J_{ij}(\bX,
        \bk) + \tilde h_i(\bX, \bk).
\end{align}
With this choice, \eqref{e:master1} will also conserve mass and have
$\mathcal{F}$ as a Lyapunov function.  In the above expression,
$\kappa$ is a rate constant, $f_i$ is the ``formation energy,'' and
$\Qm$ is the activation energy for mass exchange.\footnote{ The
  coefficient of $(f_i - f_j)$ is $\beta$ rather than
  $\frac{\beta}{2}$ because we work in the $s_i$ coordinate system.}
In principle, $\Qm$ should be obtained by finding the saddle point
associated with each pair exchange, but many practitioners take it to
be a constant for efficiency.

\section{Spin-Diffusion Model}
\label{s:spindiffusion}
\begin{figure}
  \subfigure[]{\includegraphics[width=4cm]{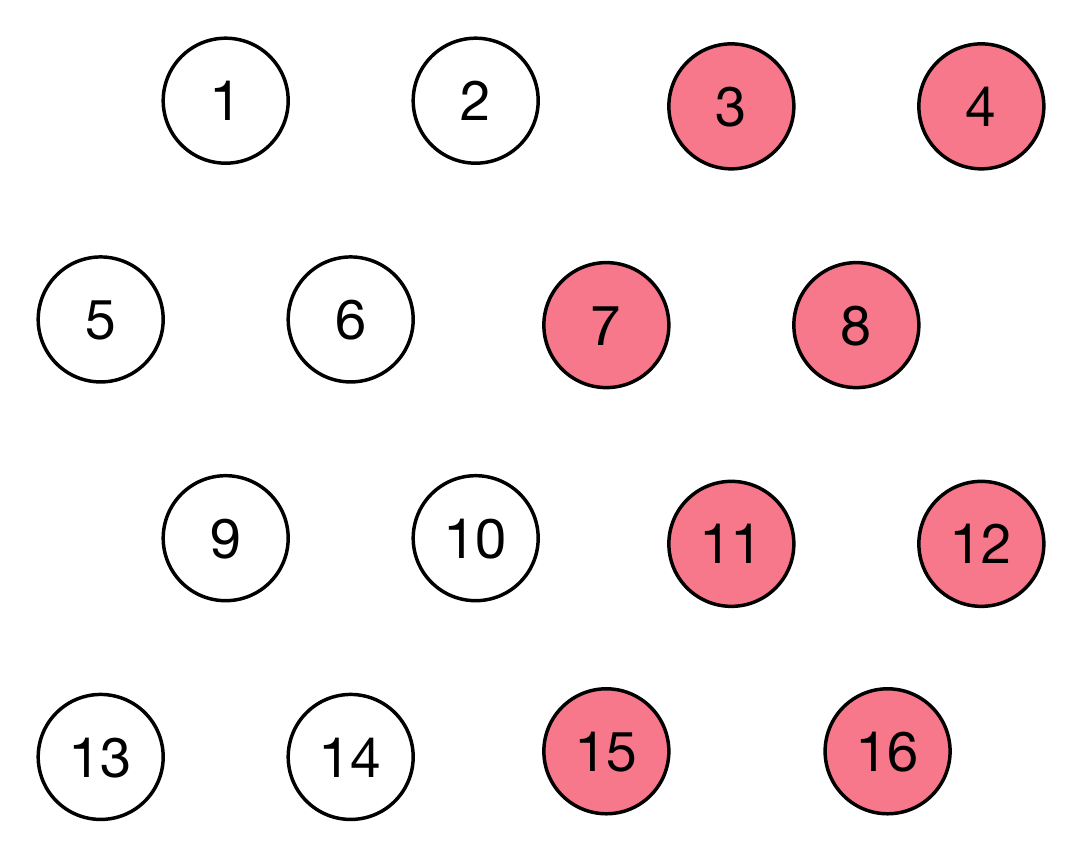}}
  \subfigure[]{\includegraphics[width=4cm]{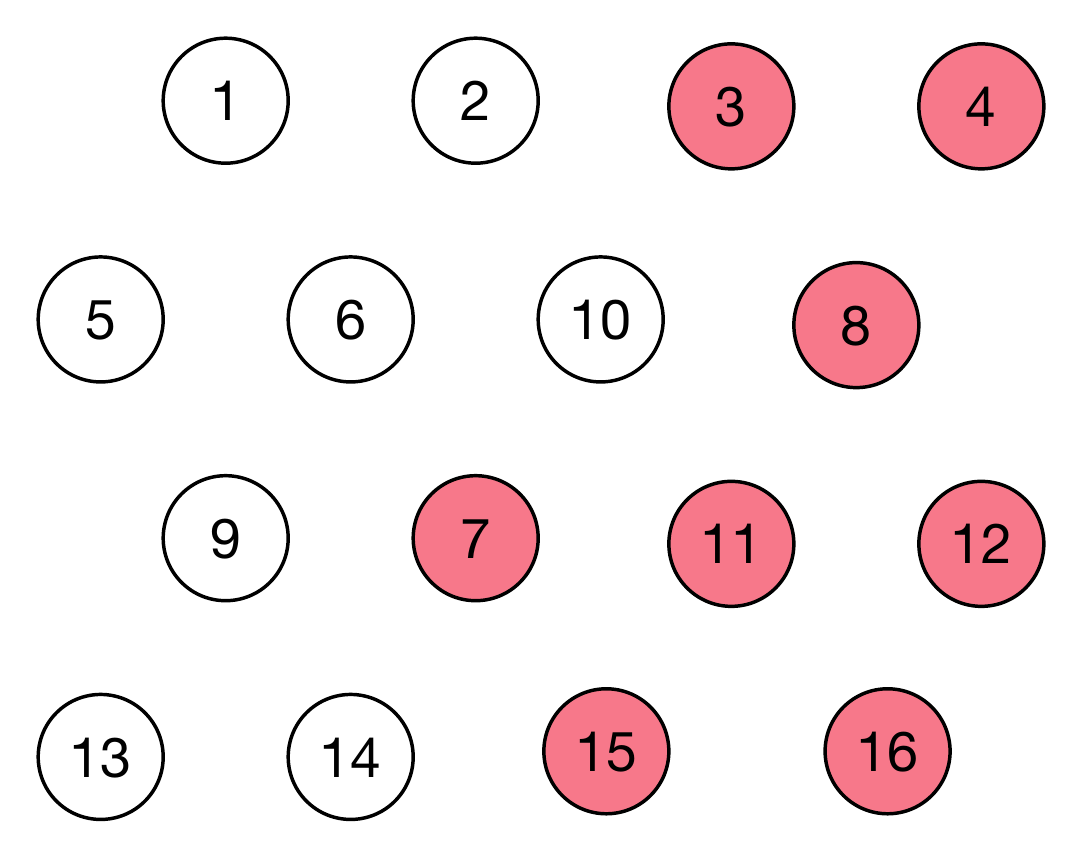}}
  \subfigure[]{\includegraphics[width=4cm]{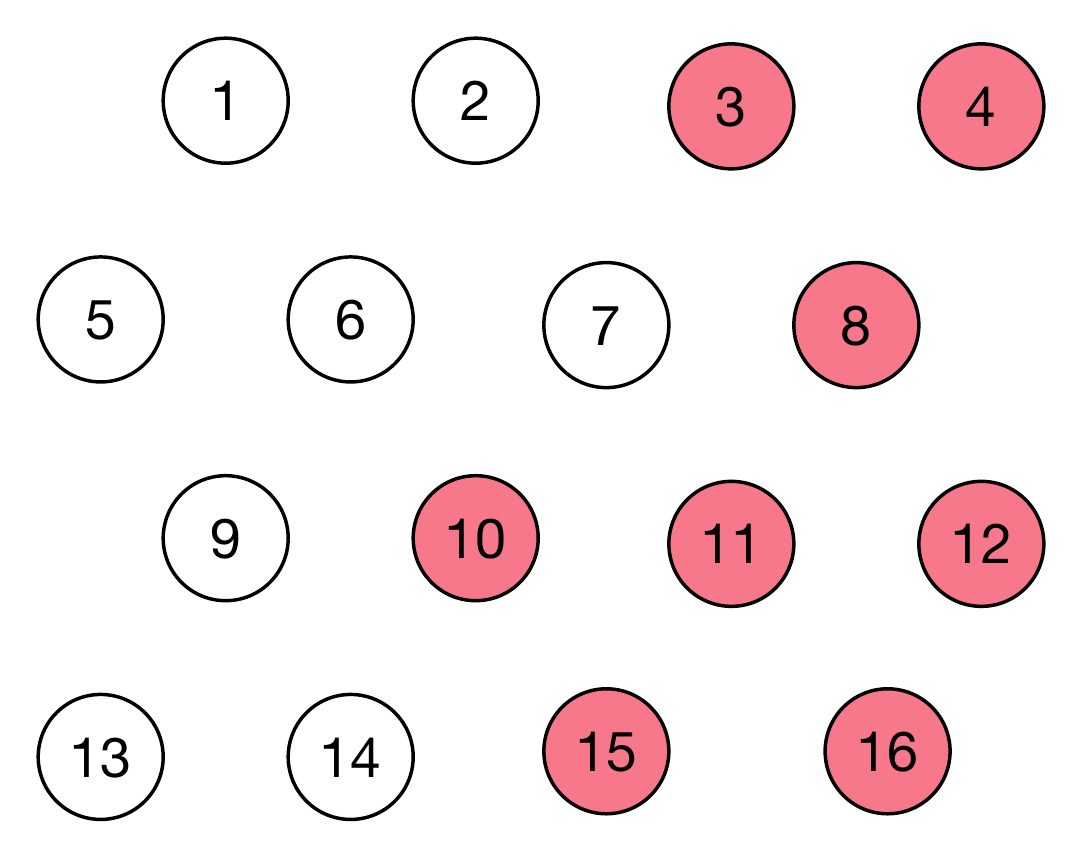}}
  \caption{Beginning in (a), we we can obtain (b) by having sites 7
    and 10 exchange with one another through the diffusion process,
    crossing multiple energy barriers along the way.  Alternatively,
    we can obtain (c) by performing a spin-exchange of the species at
    7 and 10, leaving the atomic sites in place.  {In our
      spin-diffusion model, only transitions of type (a) to (c) are
      permitted to occur.}}
  \label{f:mechanisms}
\end{figure}

{

  In this section, we develop our spin-diffusion model and show how it
  captures the essential features of DMD.  The physical intuition
  behind our model is that for the basins partitioning up the state
  space as in \eqref{e:domaindecomp}, the diffusion process will
  capture {\it intrabasin} motion while the spin-exchange process will
  capture {\it interbasin} transitions.  This idealizes the
  metastability of the basins with respect to the diffusion, assuming
  we would never observe an exit by diffusion.  Instead, the spin
  exchange migrates the system between basins.  This requires the
  introduction of the idealized {\it quasistationary distributions},
  QSDs, which characterize the distribution of diffusions conditioned
  to never leave a
  basin~\cite{lelievre2013two,Perez:2015hs,Lu:2014hl}.  We review the
  key ideas of the QSD and relate them to the spin-diffusion in the
  next subsection.  Subject to additional assumptions on the time
  scales, the reaction rates, the temperature, and a mean field
  approximation, we then obtain our evolution equations.

  The reader may question the introduction of the QSD -- why is it not
  sufficient to use the ``naive'' coupling of the free, unconditioned,
  diffusion given by \eqref{e:diffusion1} to the spin-exchange,
  \eqref{e:rates1}?  If we were to proceed with the naive coupling, as
  sketched in Section \ref{s:spinintro}, our joint process would be
  $L_\eps = \eps^{-1} L_x + L_\sigma$.  This leaves the value of the
  scale separation, $\eps$, ill-defined.  Heuristically, it is
  intended to relate the time for the diffusion to relax to a
  ``local'' equilibrium of the system, a basin of attraction
  $D_\ell(\bsigma)$ and the time for the diffusion to escape this
  basin.  But the only time scale associated with the free diffusion
  is for convergence to the Gibbs distribution on all of $\Omega^N$,
  and we wish to model the time scales of local relaxation and
  transition.  While one might attempt to use restricted Gibbs, it is
  the QSD that gives both the desired characterization of ``locality''
  and the time scales that define $\eps$.

  Additionally, the naive joint process allows events that we intend
  to restrict.  Suppose $(\bx(0), \bsigma(0))$ corresponds to Figure
  \ref{f:mechanisms} (a).  There is a nonzero chance the diffusion
  exits the basin before any spin-exchange occurs, transitioning, at
  time $t$, to Figure \ref{f:mechanisms} (b), where
  $(\bx(t), \bsigma(t)) = (\bx(0)^{7,10}, \bsigma (0)^{7,10})$. Here,
  the two atomic sites, and the their species, have exchanged through
  the diffusion.  Alternatively, a spin-exchange may occur before the
  diffusion can exit the basin, and the system may now be as in Figure
  \ref{f:mechanisms} (c), with
  $(\bx(t), \bsigma(t)) = (\bx(0), \bsigma (0)^{7,10})$.  Here, only
  the spins, but not the sites, have been exchanged.  Figures
  \ref{f:mechanisms} (b) and (c) have the same physical properties,
  but have been obtained by different mechanisms from (a). Using our
  modified joint process, where the diffusion is conditioned to never
  leave the basin, conforms to our design to only allow transitions
  via spin-exchanges, disambiguating the evolution of the system.
  This {\it conditioned diffusion} converges to a QSD.
}

\subsection{Quasistationary Distributions}
Here, we review the essential features of the QSDs for diffusions and
apply them to the spin-diffusion model.

\subsubsection{Properties}
First consider a diffusion independent of the spins,
\begin{equation}
  \label{e:diffusion2}
  d\bx(t) = -\nabla V(\bx(t)) dt + \sqrt{2\beta^{-1} }d\bW(t), \quad
  \bx(0) = \bx_0.
\end{equation}
{Excluding a set of measure zero converging to saddles, at any
  instant,} $\bx(t)$ is in a basin of attraction $D$ of the energy
$V(\bx)$, which can be identified by the local minimum to which the
system quenches:
\begin{equation}
  \label{e:localmin1}
  \lim_{\tau \to \infty} \bz(\tau;\bx), \quad\text{solving }
  \dot{\bz}=-\nabla V(\bz), \quad
  \bz(0) = \bx,
\end{equation}
and the basin is the set that quench to the same minimum:
\begin{equation}
  \label{e:basin1}
  D(\bx) = \set{\by\in \Omega^N\mid \lim_{\tau \to \infty} \bz(\tau;\by)=\lim_{\tau \to \infty} \bz(\tau;\bx)}.
\end{equation}

Given $\bx(0) = \bx_0$, we equip the generator of this process $L_x$
with Dirichlet boundary conditions on $D(\bx_0)$.  Consider the
eigenvalue problem{
  \begin{equation}
    \label{e:eval}
    \begin{split}
      L_x \varphi_i = -\nabla_x V(\bx)\cdot \nabla_x \varphi_i &+
      \beta^{-1} \Delta_x \varphi_i = -\lambda_i \varphi_i, \quad
      \varphi_i|_{\partial
        D(\bx_0)} = 0, \\
      & \int_D |\varphi_i(x)|^2 e^{-\beta V(x)}dx=1.
    \end{split}
  \end{equation}}
The eigenfunctions and eigenvalues depend parametrically on $\bx_0$
through $D=D(\bx_0)$.

We assume that $D(\bx_0)$ is metastable, which is to say that the
relaxation time scale to the QSD is far shorter than the first exit
time.  This can be quantified in terms of the first two eigenvalues:
{\begin{equation}
    \label{e:tqsd}
    t_{\rm relax} \approx \frac{1}{\lambda_2 - \lambda_1}\ll
    \frac{1}{\lambda_1}\approx t_{\rm exit},
  \end{equation}
  where we use $\approx$ to indicate these are characteristic ({\it
    i.e.} order of magnitude) time scales.}  Letting $T$ be the first
exit time, the QSD $\check \mu$ relates to the diffusion via the limit
\begin{equation}
  \label{e:qsd1}
  \lim_{t\to \infty} \P^{\bx_0}(\bx(t)\in \bullet\mid T > t) = \check \mu(\bullet;\bx_0),
\end{equation}
and it also has a density with respect to the Lebesgue measure
{\begin{equation}
    \label{e:qsddensity}
    {\check \mu}(d\bx) = \check Z^{-1} \varphi_1(\bx;\bx_0) e^{-\beta
      V(\bx)}{d\bx}, \quad \check{Z} = \int_{D(\bx_0)} \varphi_1(\bx;\bx_0) e^{-\beta
      V(\bx)}{d\bx}.
  \end{equation}}

The QSD is {not} an invariant measure for the diffusion, since
$\bx(t)$ will exit $D(\bx_0)$ in finite time a.s.  It is also {not}
the Gibbs distribution restricted to $D(\bx_0)$; by construction, the
QSD vanishes on $\partial D(\bx_0)$ and $e^{-\beta V(\bx)}$ will not.
However, in the low temperature limit, these two distributions
coincide.  The eigenfunction $\varphi_1$ in \eqref{e:qsddensity}
converges to a positive constant in the interior of the set
$D(\bx_0)$, with a boundary layer ensuring it vanishes on
$\partial D(\bx_0)$.  This has been analyzed in
\cite{DiGesu:2016tq,Lelievre:2015ii}.

\subsubsection{QSD Processes}

{As noted, a diffusion, even one whose initial position is
  sampled from the QSD, will leave the metastable region in finite
  time, a.s.  Thus, its invariant measure will not be the QSD, and
  another stochastic process which has this as its invariant measure
  is needed.}  We thus assume
\begin{assumption}[Ergodicity of a QSD Process]
  \label{asm:qsd}
  Given $\beta$ and a metastable region $D$, there exists a process
  $\check\bx(t)$ ergodic with respect to the QSD $\check \mu$.  Its
  generator, $\check L_x$ has only the constants in its kernel, while
  $\check L_x^\dag$ has only the density of the QSD, with respect to
  Lebesgue, in its kernel.
\end{assumption}

{In what follows, the choice of QSD process is not essential.
  All that is required is that a process with the QSD as its invariant
  measure and generator $\check L_x$ exists.}  An example of such a
process is a branching-interacting particle system of $M$ diffusions,
each independently obeying \eqref{e:diffusion2} until one exits
$D(\bx_0)$.  At that instant, the one which exits is killed and it is
resampled from the $M-1$ survivors; see \cite{Binder:2015gu} and
references therein.  The process, denoted $\hat\bx(t)\in D^M$ reflects
the entire ensemble,
\begin{equation}
  \label{e:ensemble}
  \hat\bx(t) = (\check\bx^{(1)}(t), \check\bx^{(2)}(t), \ldots, \check\bx^{(M)}(t)),
\end{equation}
with each $\check\bx^{(j)}(t)\sim \check\mu$, in equilibrium.  Indeed,
$\hat\bx$ has generator $\hat L_x$ with invariant measure $\hat \mu$.
We relate this back to the QSD of interest using a projection,
$\hat P$, $\check \bx(t) = \hat P(\hat \bx(t)) = \check\bx^{(1)}(t)$,
so that
\begin{equation}
  \label{e:qsdensemble}
  \E^{\check \mu}[\mathcal{O}(\bx)]= \lim_{t\to \infty}
  \E[\mathcal{O}(\hat P(\hat\bx(t)))] = \E^{\hat \mu}[\mathcal{O}(\hat P(\hat\bx))],
\end{equation}
{which is assumed to hold for all bounded, continuous observable
  functions $\mathcal{O}$.}

Quantities associated with the QSD will be indicated by $\check{}$
marks.  This is in contrast to the $\tilde{}{\;}$'s appearing on VG
averaged quantities like \eqref{e:dmdavgterms}.

\subsubsection{QSDs for the Spin-Diffusion Model}

While the basins of attraction are defined in terms of the $\bx$
argument of $V(\bx,\bsigma)$, they depend parametrically on $\bsigma$,
as in \eqref{e:domaindecomp}:
\begin{equation}
  \label{e:qsd2}
  \check\mu_{\ell} (d\bx\mid \bsigma) = \check Z_\ell (\bsigma)^{-1}
  \varphi_1(\bx; \bsigma, \ell)e^{-\beta V(\bx,\bsigma)}d\bx.
\end{equation}
$\varphi_1(\bx; \bsigma, \ell)$ is the principal eigenfunction of the
generator on the set $D_\ell (\bsigma)$.

\begin{figure}
  \centering
  \includegraphics[width=10cm]{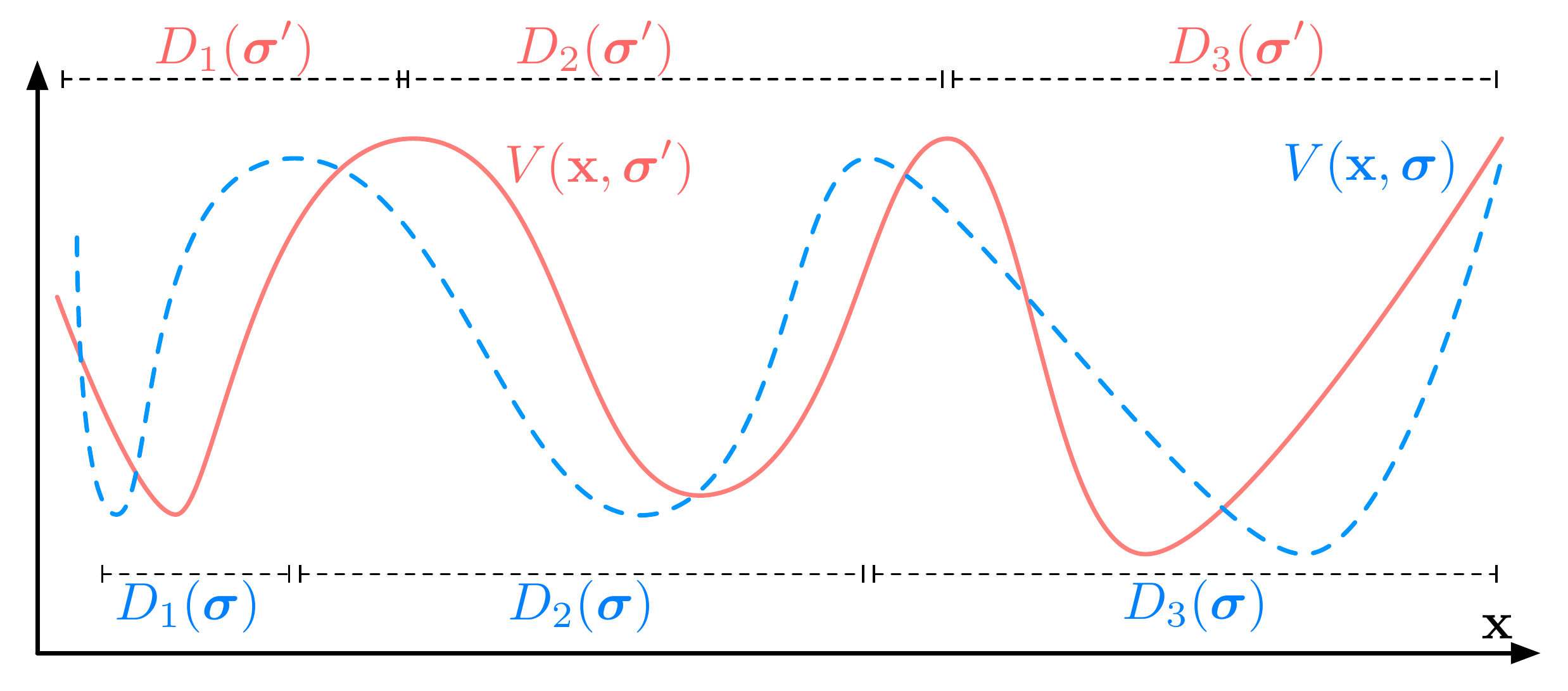}
  \caption{As the spins change, the potential energy landscape changes
    with it.  In this diagram, the system initially has spin $\bsigma$
    with $\bx \in D_2(\bsigma)$, and then the spin transitions to
    $\bsigma'$, consideration must be given as to whether it is next
    in the basin $D_1(\bsigma')$ or $D_2(\bsigma')$, since these have
    non-trivial intersection with $D_2(\bsigma)$ in the $\bx$-state
    space.}
  \label{f:Vchange}
\end{figure}

A consequence of the introduction of the QSD framework is that not
only do the spins change during the jump process,
$\bsigma\to \bsigma'$, but also the basin of attraction,
$D_\ell (\bsigma)\to D_{\ell '}(\bsigma')$.  This is due to the energy
landscape $V(\bx,\bsigma)$ changing to $V(\bx,\bsigma')$; see Figure
\ref{f:Vchange}.

The jump process is now on the joint space $(\bsigma, \ell)$.  This is
in contrast to the process sketched in the introduction, where the
reaction rates were assumed to satisfy \eqref{e:rates1}.  Instead, the
reaction rates satisfy detailed balance with respect to the QSDs: for
all $\bx \in D_\ell (\bsigma) \cap D_{\ell '}(\bsigma')$, the rates
are
\begin{equation}
  \label{e:qsdrates}
  \begin{split}
    &r((\bsigma, \ell)\to (\bsigma', \ell ');\bx)
    \varphi_1(\bx;\bsigma, \ell) e^{-\beta
      V(\bx, \bsigma)} \\
    &= r((\bsigma', \ell ')\to (\bsigma, \ell);\bx)
    \varphi_1(\bx;\bsigma', \ell ') e^{-\beta V(\bx, \bsigma')},
  \end{split}
\end{equation}
and defined to be zero outside of this intersection.  Thus, if
$ D_\ell (\bsigma) \cap D_{\ell '}(\bsigma') =\emptyset$, no exchange
is possible.  At the moment, we continue to leave the actual reaction
rates unspecified, and merely assume rates satisfying
\eqref{e:qsdrates} exist.  At fixed $\bx$, the generator of the
exchange process is
\begin{equation}
  \label{e:Lsigma}
  (\check L_{\sigma, \ell} f)(\bx,\bsigma,\ell) = \sum_{(\bsigma', \ell ')} r((\bsigma, \ell)\to (\bsigma', \ell ');\bx) [f(\bx, \bsigma',\ell') - f(\bx, \bsigma,\ell)].
\end{equation}

\subsection{Scale Separation}

The joint process for our QSD spin-diffusion has generator
\begin{equation}
  \label{e:qsdLeps}
  \check L_\eps = \eps^{-1} \check L_x + \check L_{\sigma, \ell},
\end{equation}
{We now characterize $\eps$, by assuming:}
\begin{assumption}[QSD Time Scale Separation]
  \label{asm:qsdtime}
  There is a typical value, $0<\eps\ll 1$, such that in each
  metastable set, $D_\ell(\bsigma)$,
  \begin{equation}
    \eps \approx \frac{\lambda_1(\ell, \bsigma)}{\lambda_2(\ell,
      \bsigma)-\lambda_1(\ell, \bsigma)}\ll 1,
  \end{equation}
  with $\lambda_i(\ell, \bsigma)$ the eigenvalues associated with the
  QSD on $D_\ell(\bsigma)$.
\end{assumption}\\
{Again, $\approx$ here indicates that all of these ratios are on
  the same order magnitude.}  With this choice of $\eps$, typical spin
exchanges occur on a time scale that is at least as large as a typical
diffusion first exit time from any particular basin.  Thus, our joint
process respects the physical time scales of the diffusion.

Given an observable $v(\bx, \bsigma,\ell)$, its evolution under the
process associated with $\check L_\eps$ is
\begin{equation}
  \frac{dv}{dt} = \dot v = \check L_\eps v\Rightarrow v(t) = e^{\check
    L_\eps t}v(0).
\end{equation}
{ Substituting the {\it ansatz}
  $v= v_0 + \eps v_1+\eps^2 v_2+\ldots$ into this equation leads to}
\begin{equation}
  \dot v_0 + \eps \dot v_1  + \eps^2 \dot v_2+\ldots = \eps^{-1}\check
  L_x v_0 +
  \eps^0 (\check L_x v_1 + \check L_{\sigma,\ell} v_0)+ \ldots .
\end{equation}
At $\bigo(\eps^{-1})$, we obtain
\begin{equation}
  \label{e:qsdv0}
  \check L_x v_0 = 0.
\end{equation}
Since the kernel of $\check L_x$ is assumed to be observables
independent of $\bx$, $v_0 = v_0(\bsigma, \ell,t)$.  At the next order
\begin{equation}
  \label{e:qsdv1}
  \dot v_0 = \check L_x v_1 + \check L_{\sigma, \ell} v_0 \Rightarrow \check
  L_x v_1  = \dot v_0 - \check L_{\sigma, \ell} v_0.
\end{equation}
{For the expansion to be consistent, we must satisfy the
  solvability condition,
  \begin{equation*}
    \E^{\check \mu_\ell (\bullet\mid \bsigma)}[\dot v_0 - \check
    L_{\sigma, \ell} v_0] =0.
  \end{equation*}  The leading order averaged
  equation is then}
\begin{equation}
  \label{e:qsdvavg1}
  \begin{split}
    \dot v_0 &= \E^{\check\mu_{\ell}(\bullet\mid \bsigma)}\bracket{\check L_{\sigma, \ell}} v_0\\
    &= \sum_{(\bsigma', \ell ')} \E^{\check \mu_{\ell}(\bullet\mid
      \bsigma)}[r((\bsigma, \ell)\to(\bsigma', \ell '); \bx)]
    (v_0(\bsigma', \ell ',t) - v_0(\bsigma, \ell,t) ).
  \end{split}
\end{equation}
We thus define the QSD averaged reaction rates,
\begin{equation}
  \label{e:qsdrates2}
  \check r ((\bsigma, \ell)\to(\bsigma', \ell '))  = \E^{\check \mu_{\ell}(\bullet\mid \bsigma)}[r((\bsigma, \ell)\to(\bsigma', \ell '); \bx)].
\end{equation}
These satisfy detailed balance with respect to
\begin{equation}
  \label{e:qsdavginvariant}
  \check\mu_\ell (\bsigma) = \frac{\check
    Z_\ell (\bsigma)}{\sum_{\bsigma'}\sum_{\ell '=1}^{n(\bsigma')}
    Z_{\ell '}(\bsigma')}.
\end{equation}
Therefore,
\begin{approximation}[Leading Order Process with QSDs]
  \label{apx:leadingqsd}
  Spins $\bsigma(t)$ and basins $\ell(t)$ evolve as a jump process
  with rates \eqref{e:qsdrates2}.  Atomic sites $\bx(t)$ are
  instantaneously distributed by
  $\check \mu_{\ell (t)}(\bullet\mid \bsigma(t))$.
\end{approximation}

\subsection{Master Equation}

Equation \eqref{e:qsdvavg1} and the averaged reaction rates
\eqref{e:qsdrates2} enable us to write down a master equation for the
evolution of the mean spin.  Let $p_t(\bsigma, \ell)$ be the time
dependent distribution of $(\bsigma,\ell)$.  This satisfies
\begin{equation}
  \label{e:qsdmaster1}
  \begin{split}
    \frac{d}{dt}p_t(\bsigma, \ell) &= \sum_{(\bsigma',\ell')} \check r
    ((\bsigma', \ell ')\to (\bsigma,
    \ell))p_{t}(\bsigma',\ell') \\
    &\quad - \sum_{(\bsigma',\ell')} \check r ((\bsigma, \ell)\to
    (\bsigma', \ell '))p_{t}(\bsigma,\ell).
  \end{split}
\end{equation}
Since we only allow the spin-exchanges to $\bsigma' = \bsigma^{jk}$,
with neighboring sites, $j$ and $k$, \eqref{e:qsdmaster1} becomes
\begin{equation}
  \begin{split}
    \frac{d}{dt}p_t(\bsigma, \ell) &= \sum_{\bsigma^{jk}}
    \sum_{\ell'=1}^{n} \check r ((\bsigma^{jk}, \ell')\to (\bsigma,
    \ell))p_{t}(\bsigma^{jk},\ell') \\
    &\quad -\sum_{\bsigma^{jk}} \sum_{\ell'=1}^{n} \check r ((\bsigma,
    \ell)\to (\bsigma^{jk}, \ell '))p_{t}(\bsigma,\ell).
  \end{split}
\end{equation}
Applying this to the average of spin at site $i$:
\begin{equation}
  \label{e:qsdspinmaster1}
  \begin{split}
    \frac{d}{dt}\E[\sigma_i] & =\sum_{(\bsigma,\ell)}
    \sum_{\bsigma^{jk}} \sum_{\ell'=1}^{n} \sigma_i \check r
    ((\bsigma^{jk}, \ell')\to (\bsigma,
    \ell))p_{t}(\bsigma^{jk},\ell') \\
    & \quad -\sum_{(\bsigma,\ell)} \sum_{\bsigma^{jk}}
    \sum_{\ell'=1}^{n} \sigma_i \check r ((\bsigma, \ell)\to
    (\bsigma^{jk}, \ell '))p_{t}(\bsigma,\ell).
  \end{split}
\end{equation}
Exchanging the roles of $(\bsigma, \ell)$ and $(\bsigma^{jk}, \ell')$
in the first summation of \eqref{e:qsdspinmaster1},
\begin{equation}
  \begin{split}
    &\sum_{(\bsigma,\ell)} \sum_{\bsigma^{jk}} \sum_{\ell'=1}^{n}
    \sigma_i \check r ((\bsigma^{jk}, \ell')\to (\bsigma,
    \ell))p_{t}(\bsigma^{jk},\ell') \\
    &= \sum_{(\bsigma,\ell)} \sum_{\bsigma^{jk}} \sum_{\ell'=1}^{n}
    \sigma_i^{jk} \check r ( (\bsigma, \ell)\to (\bsigma^{jk},
    \ell'))p_{t}(\bsigma,\ell).
  \end{split}
\end{equation}
Following standard treatments, such as
\cite{Kawasaki:1966kj,Kawasaki:1966wy,Penrose:1991gh,Gouyet:2003jw},
this allows us to write \eqref{e:qsdspinmaster1} as
\begin{equation}
  \label{e:qsdspinmaster2}
  \frac{d}{dt}\E[\sigma_i]   = \sum_{j\in \calN_i} \sum_{\ell' =
    1}^{n}\E[(\sigma_j - \sigma_i) \check r ( (\bsigma,
  \ell)\to (\bsigma^{ij}, \ell'))].
\end{equation}
To obtain a tractable equation, further approximations are necessary.

\subsection{Low Temperature Approximation}
{Next we make the approximation that as $\beta \to \infty$,
  $\varphi_1$ tends to a constant on $D_\ell (\bsigma)$.  Thus, in the
  low temperature regime, the QSD is approximately the restricted
  Gibbs distribution}~\cite{DiGesu:2016tq,Lelievre:2015ii}.
\begin{approximation}[Low Temperature QSD Approximation]
  \label{apx:lowtempqsd}
  $\beta$ is sufficiently large that
  \begin{equation}
    \label{e:qsdapx1}
    \begin{split}
      \check\mu_\ell(d\bx\mid\bsigma)&=\check Z_\ell(\bsigma)^{-1}
      \varphi_1(\bx;
      \bsigma, \ell) e^{-\beta V(\bx, \bsigma)}d\bx \\
      &\approx Z_\ell(\bsigma)^{-1} 1_{D_\ell (\bsigma)} e^{-\beta
        V(\bx, \bsigma)}d\bx.
    \end{split}
  \end{equation}
\end{approximation}
Hence, we can approximate \eqref{e:qsdrates} by
\begin{equation}
  \label{e:qsdratesapx}
  r((\bsigma, \ell)\to (\bsigma', \ell ');\bx) e^{-\beta
    V(\bx, \bsigma)} \\
  = r((\bsigma', \ell ')\to (\bsigma, \ell);\bx) e^{-\beta
    V(\bx, \bsigma')}
\end{equation}
for $\bx \in D_\ell (\bsigma)\cap D_{\ell '}(\bsigma')$ and zero
otherwise.  Since $\ell$ and $\ell'$ only appear in the definition of
regions $D_\ell (\bsigma)$ and $D_{\ell '}(\bsigma')$, it is easy to
formulate reaction rates satisfying this approximate detailed balance.

\subsection{State Space Approximation}

A major challenge in applying this QSD-based formulation is that we
must identify all the states neighboring $(\bsigma, \ell)$ to compute
the associated reaction rates.  While the $\bsigma'$ are easy to
identify, finding the sets $D_{\ell '}(\bsigma')$ that intersect
$D_\ell (\bsigma)$ is intractable in high dimension.

A simple approximation is to only consider a single $\ell '$ for each
$\bsigma'$, given $(\bsigma, \ell)$.  This corresponds to finding the
set $D_{\ell '}(\bsigma')$ containing the local minimum of
$D_\ell(\bsigma)$.  Consequently, if $\bsigma\to \bsigma'$, the system
will undergo a displacive relaxation from the previous local minima to
the new one.  In the context of Figure \ref{f:Vchange}, if we begin in
$D_2(\bsigma)$, then, since its local minima is in $D_2(\bsigma')$, we
go to that metastable region when $\bsigma\to \bsigma'$.  Under this
approximation given $\ell(0)$, $\ell(t)=\ell(\bsigma(t))$:
\begin{approximation}[State Space Approximation]
  \label{apx:statespace}
  Given the time series $\bsigma(t)$ and the initial basin of
  attraction indexed by $\ell(0)$, we can uniquely infer $\ell(t)$ and
  the associated basin of attraction for all time.
\end{approximation}
Thus \eqref{e:qsdspinmaster2} is replaced with
\begin{equation}
  \label{e:qsd_spinmaster3}
  \frac{d}{dt}\E[\sigma_i]   = \sum_{j\in \calN_i} \E[(\sigma_j - \sigma_i) \check r ( (\bsigma,
  \ell(\bsigma))\to (\bsigma^{ij}, \ell'(\bsigma,\bsigma^{ij})))]
\end{equation}
with the notation $\ell(\bsigma)$ and $\ell'(\bsigma, \bsigma^{ij})$
indicating that these are uniquely determined by the given spin
configuration, along with the initial conditions.

\subsection{Mean Field Approximations}

The next step is to make mean field approximations, given particular
reaction rates, to obtain a system of equations entirely in terms of
$s_i = \E[\sigma_i]$.  We consider two possible reaction rates here.
However, we emphasize that in this approximation we are not taking the
large volume, fixed density, limit; discreteness of the system is
retained.

\subsubsection{$\tanh$ Reaction Rates}

A simple choice of reaction rate satisfying \eqref{e:qsdratesapx}
involves the $\tanh$ function:
\begin{equation}
  \label{e:tanh1}
  r(\bsigma\to \bsigma^{ij}; \bx) = \tau^{-1}\set{\tfrac{1}{2} -
    \tfrac{1}{2}\tanh\paren{\tfrac{\beta}{2}\Delta_{ij} V(\bx,\bsigma)}},
\end{equation}
with $\Delta_{ij} V(\bx,\bsigma)$ given in \eqref{e:deltaV}, and
$\tau$ a rate constant.  Note that through the introduction of $\eps$
in \eqref{e:qsdLeps}, after integrating out to time
$t=\bigo(\eps\lambda_1^{-1})= \bigo((\lambda_2-\lambda_1)^{-1})$ an
exit event would have occurred in the underlying diffusion, and now a
spin-exchange should occur in the joint process.  Hence,
$\tau = \bigo((\lambda_2-\lambda_1)^{-1})$.  After the scale
separation, units of time and $\tau$ can be arbitrarily rescaled,
jointly.

Following \cite{Penrose:1991gh}, the conditional average is:
\begin{equation}
  \label{e:tanhrate1}
  \begin{split}
    & (\sigma_j - \sigma_i)\E^{\check\mu_\ell(\bullet\mid\bsigma)}[ r(\bsigma\to \bsigma^{ij};\bx)]  \\
    & = \frac{1}{2}\tau^{-1} (\sigma_j - \sigma_i)-
    \frac{1}{2}\tau^{-1} (1- \sigma_i\sigma_j)
    \E^{\check\mu_\ell(\bullet\mid\bsigma)}\bracket{\tanh\paren{{\beta}\frac{\Delta_{ij}
          V(\bx,\bsigma)}{\sigma_j-\sigma_i}}}.
  \end{split}
\end{equation}
No approximation has yet been made in \eqref{e:tanhrate1}.  To obtain
equations of motion entirely in terms of $\bs(t)$, we now make:
\begin{approximation}[Mean Field Approximation for $\tanh$ Reaction
  Rates]
  \label{apx:qsd_mf1}
  Assume that:
  \begin{itemize}
  \item The averages over $\check\mu_\ell(\bullet\mid \bsigma)$
    approximately commute with the $\tanh$ function.
  \item $\check\mu_\ell(\bullet\mid \bsigma)$ can be approximated by
    $\check\mu_\ell(\bullet\mid \bs)$.
  \item Spins are approximately uncorrelated,
    $\E[\sigma_i\sigma_j]\approx \E[\sigma_i]\E[\sigma_j]$.
  \end{itemize}
  Then \eqref{e:qsd_spinmaster3} can be approximated by the system
  \begin{equation}
    \label{e:qsd_mfmaster1}
    \frac{d}{dt}s_i = \frac{1}{2}\tau^{-1}\sum_{j \in \calN_i} (s_j - s_i) -
    (1-s_is_j) \tanh\paren{\beta \frac{\Delta_{ij}\check V^{(\ell(\bs))}(\bs)}{s_j-s_i}}.
  \end{equation}
  with atomic sites
  $\bx(t) \sim \check\mu_{\ell(\bs)}(\bullet\mid \bs)$.
\end{approximation}\\
{The QSDs $\check \mu_{\ell}(\bs)$ are defined analogously to
  $\check \mu_{\ell}(\bsigma)$ with $s_i$ taking the place of
  $\sigma_i$ in the potential in \eqref{e:modelV2}. The basins are now
  defined as $D_\ell(\bs)$, which are implicitly assumed to still
  satisfy a corresponding version of Assumption \ref{asm:finiten}.}
The argument of the $\tanh$ function in \eqref{e:qsd_mfmaster1} is
\begin{equation}
  \label{e:qsdtanh1}
  \begin{split}
    \frac{\Delta_{ij} \check V^{(\ell(\bs))} (\bs)}{s_j-s_i} =
    {-\sum_{k\neq i,j} (\check J_{ik}^{(\ell(\bs))}(\bs) - \check
      J_{jk}^{(\ell(\bs))} (\bs ) )s_k + (\check h_i^{(\ell(\bs))}
      (\bs) - \check h_j^{(\ell(\bs))}(\bs))}
  \end{split}
\end{equation}
with
\begin{equation}
  \label{e:qsdtanhavg1}
  \begin{split}
    \check J_{ij}^{(\ell(\bs))}(\bs) =
    \E^{\check\mu_{\ell(\bs)}(\bullet\mid\bsigma)}[J_{ij}(\bx)],\quad
    \check h_{i}^{\ell(\bs)} (\bs)= \E^{\check\mu_\ell
      (\bullet\mid\bs)}[h_{i}(\bx)], \\
    \Delta_{ij}\check V^{(\ell(\bs))} (\bs) =
    \E^{\check\mu_{\ell(\bs)} (\bullet\mid\bs)}[\Delta_{ij} V(\bx,
    \bs))].
  \end{split}
\end{equation}
These are the QSD averaged analogs of the VG averages in
\eqref{e:dmdavgterms}.  The symbol ${\ell(\bs)}$ is used in the above
expressions to remind the reader that we have made Approximation
\ref{apx:statespace} on the state space.  For brevity, we will
suppress it in what follows.

\subsubsection{Arrhenius Reaction Rates}
Another choice is the Arrhenius type reaction rates, subject to a
constant saddle height approximation:
\begin{equation}
  \label{e:arrhenius1}
  r(\bsigma\to \bsigma^{ij}; \bx) = \tau^{-1}\exp\paren{-\tfrac{\beta}{2}\Delta_{ij}V(\bx, \bsigma)}.
\end{equation}
In analogous fashion to Approximation \ref{apx:qsd_mf1},
\begin{approximation}[Mean Field Approximation for Arrhenius Reaction
  Rates]
  \label{apx:qsd_mf2}
  Assume that:
  \begin{itemize}
  \item The averages over $\check\mu_\ell(\bullet\mid \bsigma)$
    approximately commute with the $\tanh$ function.
  \item $\check\mu_\ell(\bullet\mid \bsigma)$ can be approximated by
    $\check\mu_\ell(\bullet\mid \bs)$.
  \item Spins are approximately uncorrelated,
    $\E[\sigma_i\sigma_j]\approx \E[\sigma_i]\E[\sigma_j]$.
  \end{itemize}
  Then \eqref{e:qsd_spinmaster3} can be approximated by the system
  \begin{equation}
    \label{e:qsd_mfmaster2}
    \begin{split}
      \frac{d}{dt}s_i = &\sum_{j\in \calN_i} \frac{1}{2}\tau^{-1} (1-s_i)(1+s_j) \exp\paren{-\beta\frac{\Delta_{ij}\check  V(\bs)}{s_j-s_i}}\\
      &\quad - \frac{1}{2}\tau^{-1} (1+s_i)(1-s_j) \exp\paren{\beta
        \frac{\Delta_{ij}\check V(\bs)}{s_j-s_i}}.
    \end{split}
  \end{equation}
  with atomic sites
  $\bx(t) \sim \check\mu_{\ell(\bs)}(\bullet\mid \bs)$.
\end{approximation}

\subsection{Comparison with DMD}
\label{s:comp1}
In this section we compare the models \eqref{e:qsd_mfmaster1} and
\eqref{e:qsd_mfmaster2}, with the DMD models \eqref{e:gradientflow1}
and \eqref{e:master1}.  Our models are close in spirit to DMD
dynamics, with the significant difference being the use of terms like
$\tilde J_{ij}^{(\ell)}(\bs, \bX, \bk)$ from~\eqref{e:dmdavgterms} in
place of $\check J_{ij}^{(\ell)}(\bs)$.  This reflects that, in
standard DMD models, a VG approximation of a single mode of the Gibbs
distribution is used, while we use a QSD.  Both distributions capture
local features of $e^{-\beta V(\bx,\bsigma)}$.  {Furthermore, in
  the low temperature limit, the QSD will be well approximated by
  restricted Gibbs (see Approximation \ref{apx:lowtempqsd}), and
  individual modes of $e^{-\beta V(\bx,\bsigma)}$, will be well
  approximated by restricted Gaussians ({\it i.e.}, with harmonic
  potentials).  Thus, we expect agreement of terms like
  $\tilde J_{ij}^{(\ell)}(\bs, \bX, \bk)$ and
  $\check J_{ij}^{(\ell)}(\bs)$ in this limit.}

\subsubsection{$\tanh$ Reaction Rates}

For the $\tanh$ reaction rate model, suppose we make a high
temperature approximation of \eqref{e:qsd_mfmaster1} to expand the
$\tanh$ function
\begin{equation}
  \label{e:mfmaster1hightemp}
  \frac{d}{dt}s_i  =\frac{1}{2}\tau^{-1}\sum_{j \in \calN_ i} (s_j - s_i) -
  (1-s_is_j){\beta \frac{\Delta_{ij}
      \check V(\bs)}{s_j-s_i}} + \bigo(\beta^2).
\end{equation}
Next, we use \eqref{e:Fgrad} and \eqref{e:deltaV} to write the
difference in the free energy gradients as
\begin{equation}
  \label{e:Fgraddiff}
  \begin{split}
    \frac{\partial\mathcal{F}}{\partial s_j }-\frac{\partial
      \mathcal{F}}{\partial s_i} &= -\frac{\Delta_{ij} \tilde V(\bs,
      \bX, \bk)}{s_j - s_i}+ \tilde
    J_{ij} (s_j - s_i) \\
    &\quad + \beta^{-1} \arctanh(s_j)-\beta^{-1} \arctanh(s_i).
  \end{split}
\end{equation}
Under a constant mobility model, we can write \eqref{e:gradientflow1}
as
\begin{equation}
  \label{e:gradientflow2}
  \begin{split}
    \frac{d}{dt}s_i =\beta^{-1} m_{\rm c} &\sum_{j\in
      \calN_i}\arctanh(s_j)- \arctanh(s_i)
    -\beta\frac{\Delta_{ij} \tilde V(\bs, \bX, \bk)}{s_j - s_i}\\
    &\quad \quad + \beta \tilde J_{ij} (s_j - s_i).
  \end{split}
\end{equation}
Next, neglecting boundary conditions, in the high temperature, near
equilibrium limit, we expect, from rigid lattice asymptotics,
$s_i = M/N + \bigo(\beta) = s_{\eq} + \bigo(\beta)$, so that
\eqref{e:mfmaster1hightemp} becomes
\begin{equation}
  \label{e:mfmaster1hightemp2}
  \frac{d}{dt}s_i = \frac{1}{2}\tau^{-1}\sum_{j \in \calN_ i} (s_j
  -s_i)-\beta (1-s_{\eq}^2) \frac{\Delta_{ij}
    \check V(\bs)}{s_j-s_i} + \bigo(\beta^2)
\end{equation}
while \eqref{e:gradientflow2} becomes
\begin{equation}
  \label{e:gradientflow3}
  \begin{split}
    \frac{d}{dt}s_i &=\beta^{-1} m_{\rm c} \sum_{j\in \calN_i}
    \frac{s_j - s_i}{1-s_{\eq}^2} - \beta\frac{\Delta_{ij} \tilde
      V(\bs, \bX, \bk)}{s_j - s_i}+ \beta\tilde J_{ij} (s_j - s_i).
  \end{split}
\end{equation}

Comparing \eqref{e:gradientflow2} with \eqref{e:mfmaster1hightemp2},
we see what further conditions must hold for the two models to be
approximately equivalent.  First, the VG approximation must be
suitable,
\begin{equation}
  \label{e:comp1}
  \check V(\bs) \approx \tilde V(\bs,\bX,\bk),\quad \check V(\bs^{\ij})
  \approx \tilde V(\bs^{\ij},\bX,\bk).
\end{equation}
Next, there is also the additional term,
$\beta \tilde J_{ij} (s_j - s_i) $, at $\bigo(\beta)$ which was
discussed in \cite{Penrose:1991gh}, in the case of a rigid lattice.
In that work, the author commented that such terms motivated the use
of the mean field kinetic models, in the spirit of
\eqref{e:qsd_mfmaster1}, rather than the free energy gradient model,
\eqref{e:gradientflow1}.  Finally, the mobility should scale as
\begin{equation}
  \label{e:kcscaling}
  m_{\rm c} \approx \tfrac{1}{2}\beta\tau^{-1}(1-s_{\eq}^2),
\end{equation}
which is consistent with \cite{Sarkar:2011wu}. A similar analysis
holds the rate limited mobility model, where, in the high temperature
near equilibrium limit,
\[
  m_{ij} = m_{\rm r}[ (1- s_{\eq}^2) + \bigo(\beta)]
\]
resulting in
\begin{equation}
  \label{e:krscaling}
  m_{\rm r} \approx \tfrac{1}{2}\beta\tau^{-1}
\end{equation}

A serious weakness of this analysis is the use of high temperature
approximations.  This specifically voids Approximation
\ref{apx:lowtempqsd}, and it degrades the quality of the VG
approximation.  Nevertheless, it provides some basic insight into how
the models compare, and offers rough scalings for constants like
\eqref{e:kcscaling} and \eqref{e:krscaling}.

\subsubsection{Arrhenius Reaction Rates}

For the Arrhenius reaction rate model, we substitute identity
\eqref{e:Fgraddiff} into \eqref{e:jumprates1}
and~\eqref{e:formationenergy1} to express the DMD model
\eqref{e:master1} as
\begin{equation}
  \label{e:master1_v2}
  \begin{split}
    \frac{d}{dt}s_i= &\sum_{j\in \calN_i} \kappa e^{-\beta \Qm}
    (1+s_j)(1-s_i)\exp\paren{-\beta \frac{\Delta_{ij}\tilde
        V}{s_j-s_i}
      + \beta \tilde J_{ij} (s_j-s_i)} \\
    & \quad - \kappa e^{-\beta \Qm} (1-s_j)(1+s_i) \exp\paren{\beta
      \frac{\Delta_{ij}\tilde V}{s_j-s_i} - \beta \tilde J_{ij}
      (s_j-s_i)}.
  \end{split}
\end{equation}
Comparing with \eqref{e:qsd_mfmaster2}, we see a distinction in the
multiplicative terms
$\exp(\pm \beta \tilde J_{ij} (s_j-s_i)) = 1 + \bigo(\beta)$.  Again,
this corresponds to the term identified in \cite{Penrose:1991gh}.  We
thus infer
\begin{equation}
  \kappa e^{-\beta Q_{\rm m}}\approx \tfrac{1}{2} \tau^{-1},
\end{equation}
provided \eqref{e:comp1} holds. Models \eqref{e:master1} and
\eqref{e:qsd_mfmaster2} have similar equations, and there is no need
for an expansion in $\beta$, excepting the term
$\exp(\pm \beta \tilde J_{ij} (s_j-s_i))$.  In the high temperature
limit, all of these models tend to agree, to leading order.

\section{Numerical Simulations}
\label{s:numerics}

Here, we compare some of the models presented in this paper with a
simple test problem.

\subsection{Test Problem}
\label{s:test}

\begin{figure}
  \centering
  \includegraphics[width=10cm]{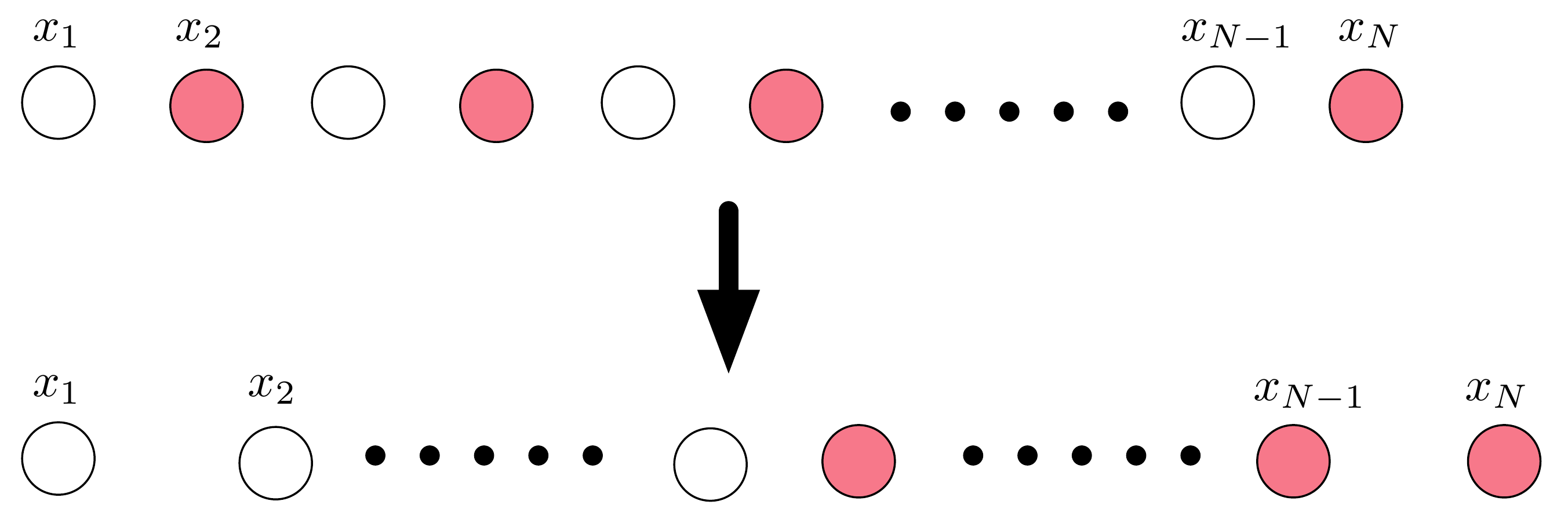}
  \caption{A chain of two species, initially arranged in an
    alternating sequence.  Segregation into two components is
    energetically favorable and leads to extension due to the choice
    of parameters in the potentials.}
  \label{f:chain_diag}
\end{figure}

Our test problem is a 1D binary alloy, consisting of a chain of atomic
sites that are occupied either by Species A or Species B. The pair
potential $\phi_{\sigma_i,\sigma_j}(|x_{ij}|)$ is such that is
energetically favorable to have like species neighbor one another (A-A
or B-B bonds).  Additionally, A-A and B-B bonds have longer
equilibrium interatomic spacings than A-B bonds resulting in
mechanical deformation in response to rearrangement of the species.
See Figure \ref{f:chain_diag} for the problem setup. While this
problem has certain nonphysical aspects, it captures the essential
features of our spin-diffusion model.  This allows for comparisons
with a number of deterministic models, including DMD free energy
gradient dynamics and mean field master equations.

A few additional details on this test problem are of note.  Owing to
the translation invariance of the problem, the leftmost atom is pinned
at $x_1 =0$, while the rightmost atom is free to move.  The species of
the atoms at the first and last atomistic sites are fixed, and only
$s_i$, $i=2,\ldots, N-1$, evolve under the flow.

With respect to energy model \eqref{e:modelV1}, there is no
$u_{\sigma}$ contribution, and the pair potential is given by the
smoothed Lennard-Jones type potential \eqref{e:ljsmooth} with cutoff
\eqref{e:cutoff}.  The variations to the pair potentials by bond type
are controlled by the choice of parameters.  We also add the confining
pair potential \eqref{e:confine} between adjacent atoms to ensure that
there is no disaggregation of the chain.

\subsection{Model Comparison}

As discussed throughout this work, there are many choices of dynamics.
Here, we compare the following:
\begin{itemize}
  {\item The spin-diffusion model, with scale separation, low
    temperature approximation, and the state space approximation,
    corresponding to \eqref{e:qsd_spinmaster3} with rates
    \eqref{e:tanh1}.}

\item Free energy gradient dynamics \eqref{e:gradientflow1}, with
  constant mobility model \eqref{e:constmobility}.
\item Free energy gradient dynamics \eqref{e:gradientflow1}, with
  mobilities defined by \eqref{e:sarkarmobility}.
\item Mean field Approximation \ref{apx:qsd_mf1}, given by equation
  \eqref{e:qsd_mfmaster1}, with an additional VG approximation of
  $\Delta_{ij} \check V(\bs)$, as in \eqref{e:comp1}.
\item Mean field Approximation \ref{apx:qsd_mf1}, given by equation
  \eqref{e:qsd_mfmaster1}, with an additional point estimate of
  $\Delta_{ij} \check V(\bs)$:
  \begin{equation}
    \label{e:point_apx1}
    \Delta_{ij} \check V (\bs) = \E^{\check\mu_\ell(\bullet\mid\bs)}\bracket{\Delta_{ij}
      V(\bx,\bs)}
    \approx{\Delta_{ij}
      V(\bX_{\min},\bs)},
  \end{equation}
  with $(\bX_{\min}, \bk_{\min})$ the local minimizer of
  $\mathcal{F} (\bs, \bX, \bk)$.
\end{itemize}

To fix a time scale, we choose $\tau = 1$
in~\eqref{e:qsd_mfmaster1}. The free energy gradient dynamics with
mobilities~\eqref{e:constmobility} and~\eqref{e:sarkarmobility} have
free parameters $m_{\rm c}$ and $m_{\rm r}$, respectively. These are
chosen so that the mean spins evolve on a similar time scales in both
the mean field and free energy gradient models.  While we calculated
scalings \eqref{e:kcscaling} and \eqref{e:krscaling} in the
high-temperature, near-equilibrium limits, they might not extend to
low temperatures.

\begin{figure}
  \centering
  \subfigure[$N=32$]{\includegraphics[width=6.25cm]{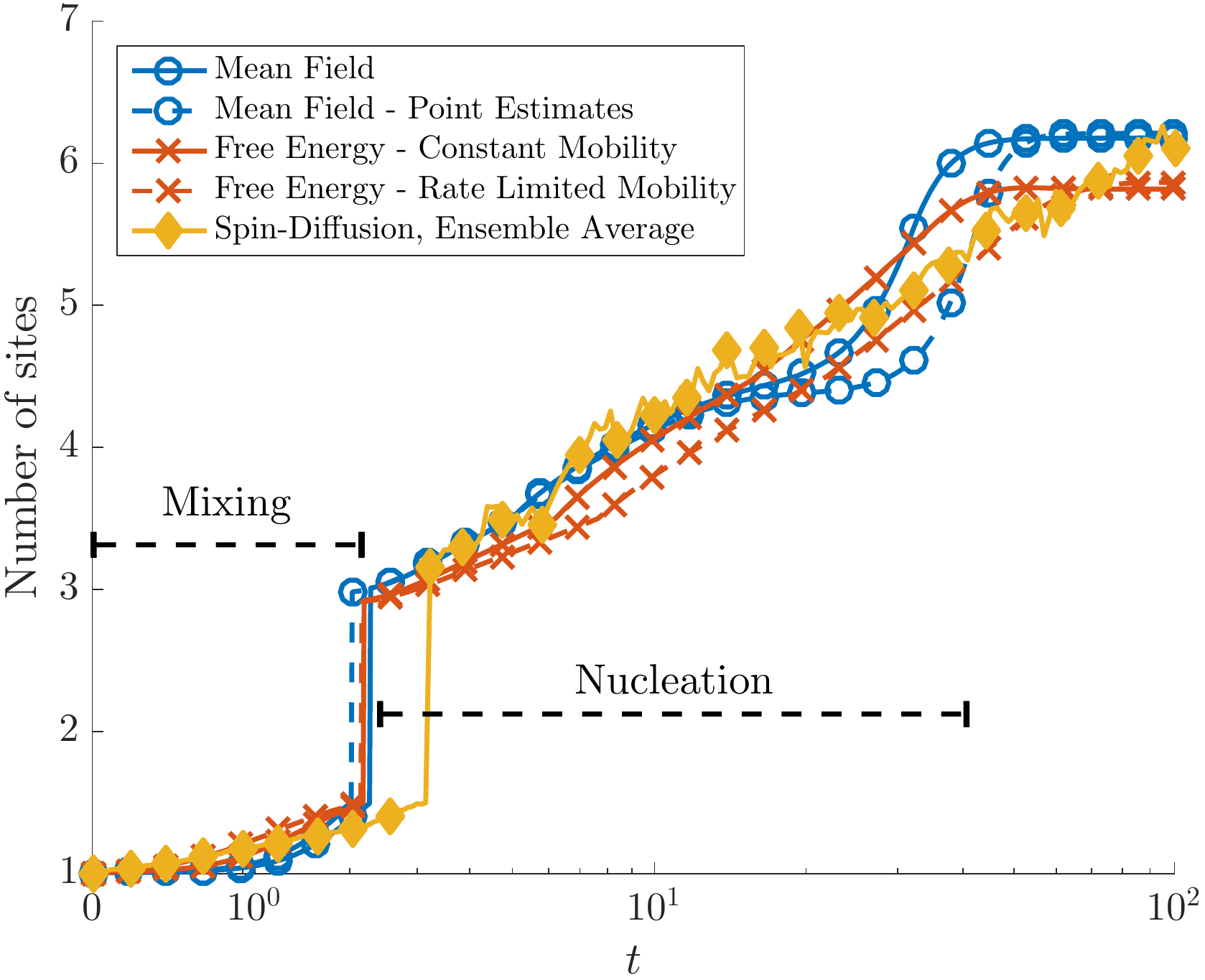}}
  \subfigure[$N=500$]{\includegraphics[width=6.25cm]{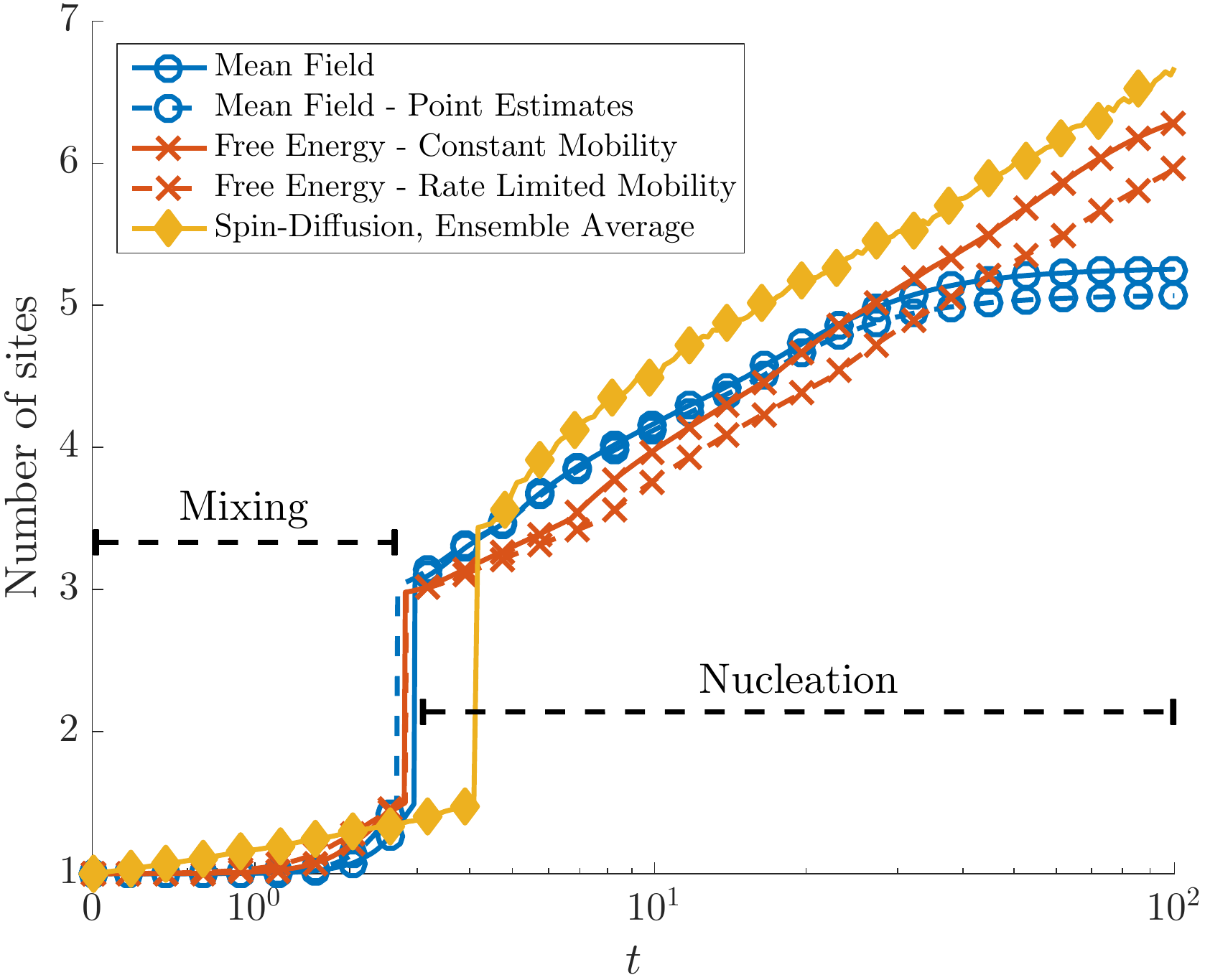}}

  \caption{Location of the first minimum in the spatial spin
    autocorrelation, \eqref{e:Rauto} at $\beta = 160$.  Simulations
    with $N=1000$ and $N=2000$ systems were in close agreement with
    $N=500$. }

  \label{f:N32beta160scaling}
\end{figure}

Indeed, we choose $m_{\rm c}$ and $m_{\rm r}$ empirically to align a
reference feature of the dynamics across the different models.  Here,
this feature corresponds to the first jump in the minimum of the
spatial spin autocorrelation, {
  \begin{equation}
    \label{e:Rauto}
    R(j) = \frac{1}{N-j} \sum_{i=1}^{N-j} (s_i - \bar{\bs})(s_{i+j} -
    \bar{\bs}), \quad \bar{\bs} = \frac{1}{N}\sum_{i=1}^N s_i.
  \end{equation}
  The first minimum of $R(j)$ corresponds to the characteristic width
  of a spatial domain.  The first jump in this minimizer, estimated as
  a continuous quantity via interpolation, is shown in Figure
  \ref{f:N32beta160scaling}.  This is further discussed below.}  After
running the simulations with $m_{\rm c}=4$ and $m_{\rm r} = 1$, time
is rescaled to match the reference feature.  While this fitting is
done independently for each $N$ , the values in Table
\ref{t:mobilities_beta160} reveal consistency across the values.

\begin{table}
  \centering
  \caption{Effective mobility for free energy gradient dynamics with
    $\beta=160$.}
  \label{t:mobilities_beta160}

  \begin{tabular}{lcrrrr}
    \hline
    \hline
    $N$ 	&  $\tfrac{1}{2}\beta\tau^{-1}$ scaling&	32 & 500 &1000 &2000\\
    \hline
    $m_{\rm c}$ &	80 &52 &	48 &47 &	46\\
    $m_{\rm r}$ &	80 &37 &	36 &36 &	36\\
    \hline
  \end{tabular}
\end{table}

\subsection{Numerical Implementation}

{For initial conditions, the $\sigma_i$ of the stochastic model
  alternate $\pm 1 $, while in the deterministic models, the $s_i$
  alternate $\pm 0.9998$.  Given this initial spin composition, the
  initial conditions in the stochastic model are obtained by quenching
  the energy to a minimizer, while $\bX$ and $\bk$ are chosen to
  locally minimize the free energy via gradient descent. }

{Our simulation of \eqref{e:qsd_spinmaster3} is performed by
  integrating the overdamped Langevin until it samples the QSD, after
  which a spin-exchange is performed using the time averaged reaction
  rates.  The diffusion is integrated using preconditioned MALA. }

{In the deterministic models, the $\bs$ are evolved by one of the
  choices of dynamics, and, since each model involves a VG
  approximation, the $\bX$ and $\bk$ are evolved by a scaled gradient
  descent for the free energy.} Thus, the system that is actually
solved is:
\begin{subequations}
  \label{e:matlabODE}
  \begin{align}
    \dot \bs &= f(\bs, \bX, \bk) \label{e:ds_dt}\\
    \dot \bX&= - \eta_1 \nabla_{\bX} \mathcal{F} (\bs, \bX, \bk) \label{e:dX_dt}\\
    \dot \bk&= -\eta_2 \nabla_{\bk} \mathcal{F} (\bs, \bX, \bk).\label{e:dk_dt}
  \end{align}
\end{subequations}
Here, $f$ comes from one of the models mentioned above.  The constants
$\eta_1$ and $\eta_2$ are large numerical parameters that force the
system to be close to the local minimizer of
$\mathcal{F}(\bs, \bX, \bk)$.

This system of ODEs is solved in {\sc Matlab} using the {\tt ode15s}
stiff solver.  If the magnitude of $\nabla_{\bX,\bk}\mathcal{F}$
exceeds a threshold tolerance, the integrator is halted and gradient
descent is applied to obtain a local minimizer, again.  After
minimization, integration of \eqref{e:matlabODE} resumes.  For
efficiency, quadrature is performed using the GSL implementation of
QUADPACK routines, \url{http://www.gnu.org/software/gsl/}, while
gradient descent minimization is done by running {\sc Matlab}'s {\tt
  ode15s} on the gradient dynamics to steady state.

Additional details on our numerical schemes are given in
{Appendices~\ref{a:numerics} and~\ref{a:numerics_spinDiff}}.

\subsection{Results}

Here, we present results of the simulation for a small system with
$N=32$, allowing us to observe near complete coarsening, along with
larger systems with $N=500$, $1000$ and $2000$, to see more physically
meaningful behavior.

\subsubsection{Small System}

\begin{figure}
  \centering \subfigure[Single
  Realization]{\includegraphics[width=6.25cm]{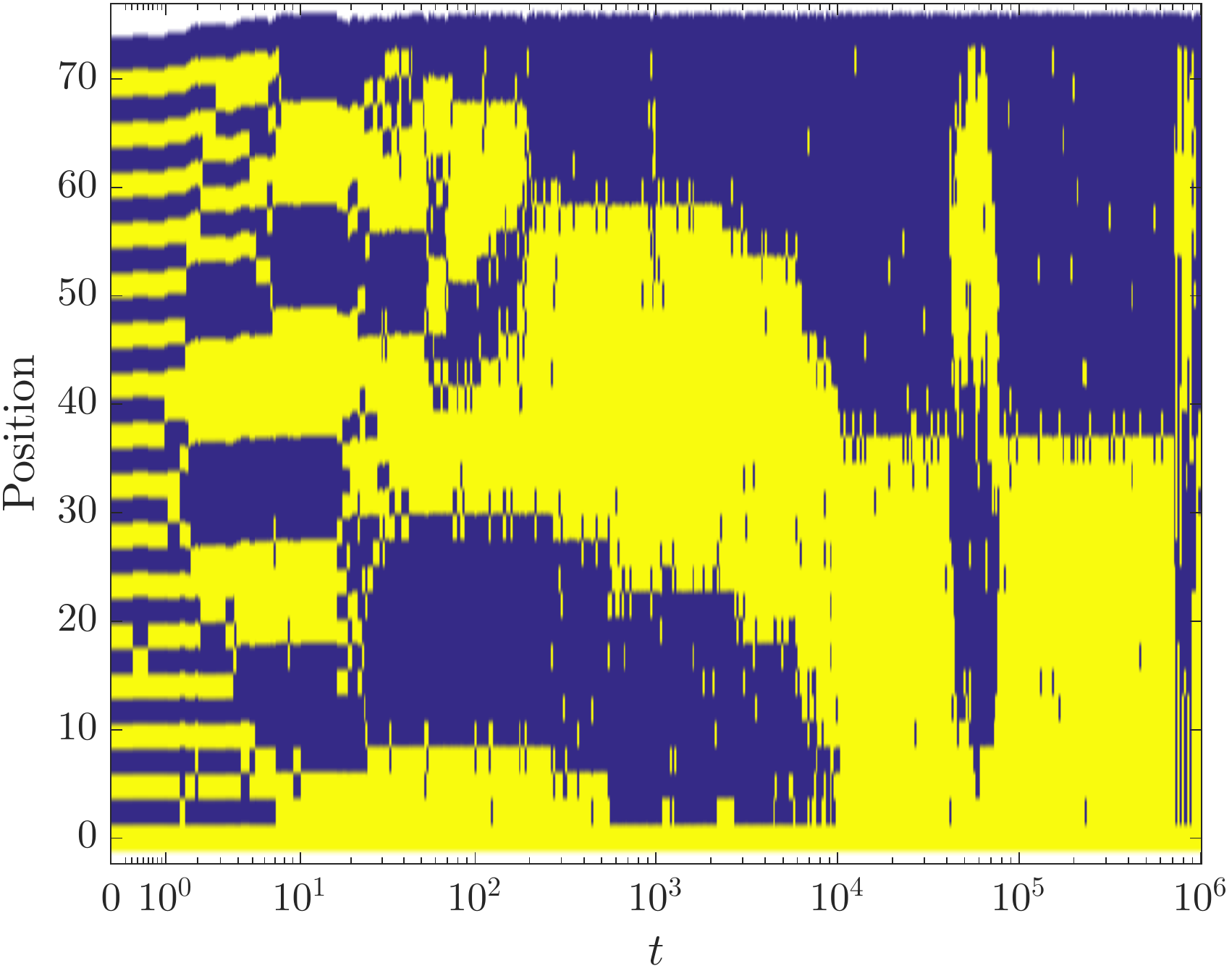}}
  \subfigure[Ensemble
  Average]{\includegraphics[width=6.25cm]{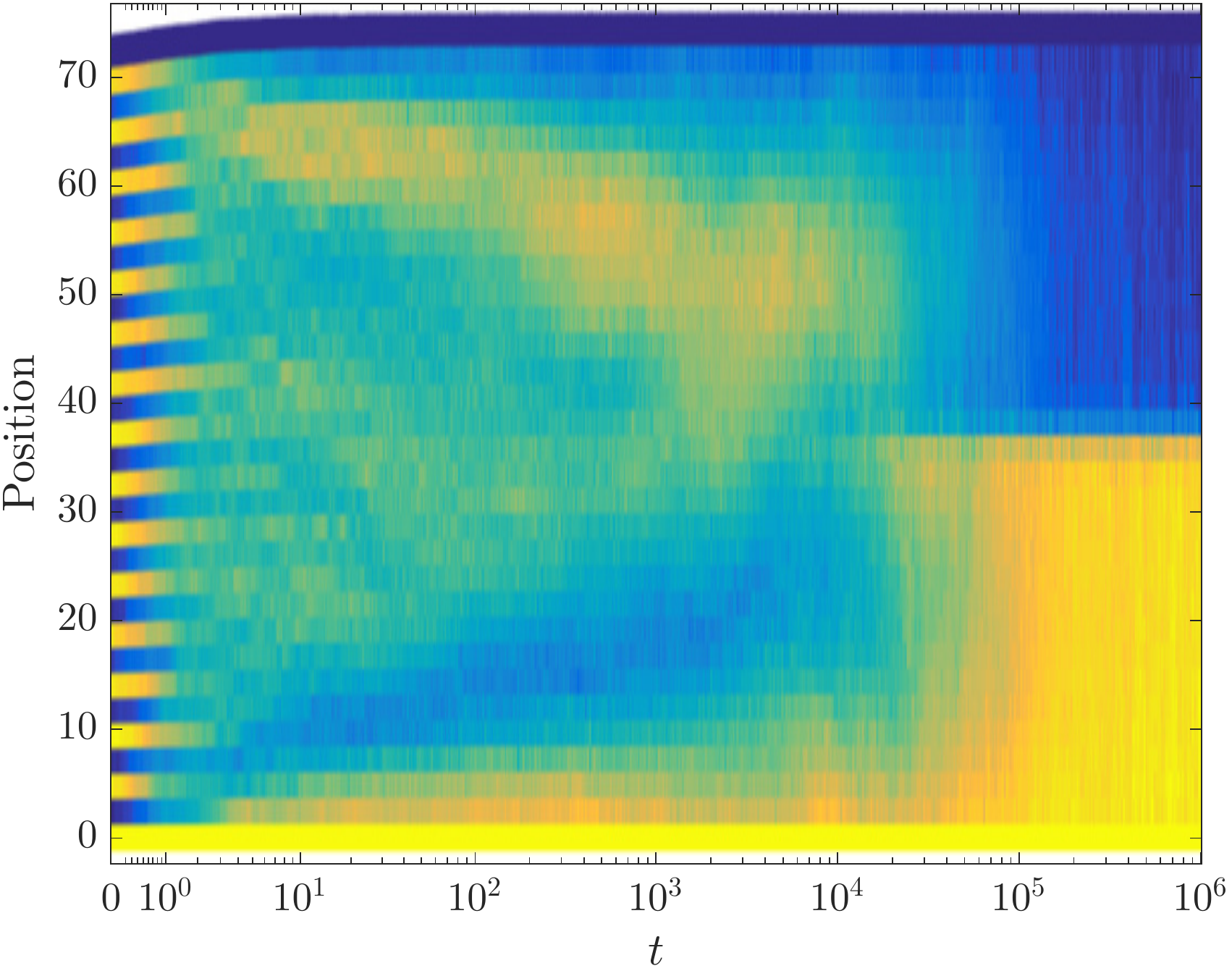}}

  \subfigure[Mean
  Field]{\includegraphics[width=6.25cm]{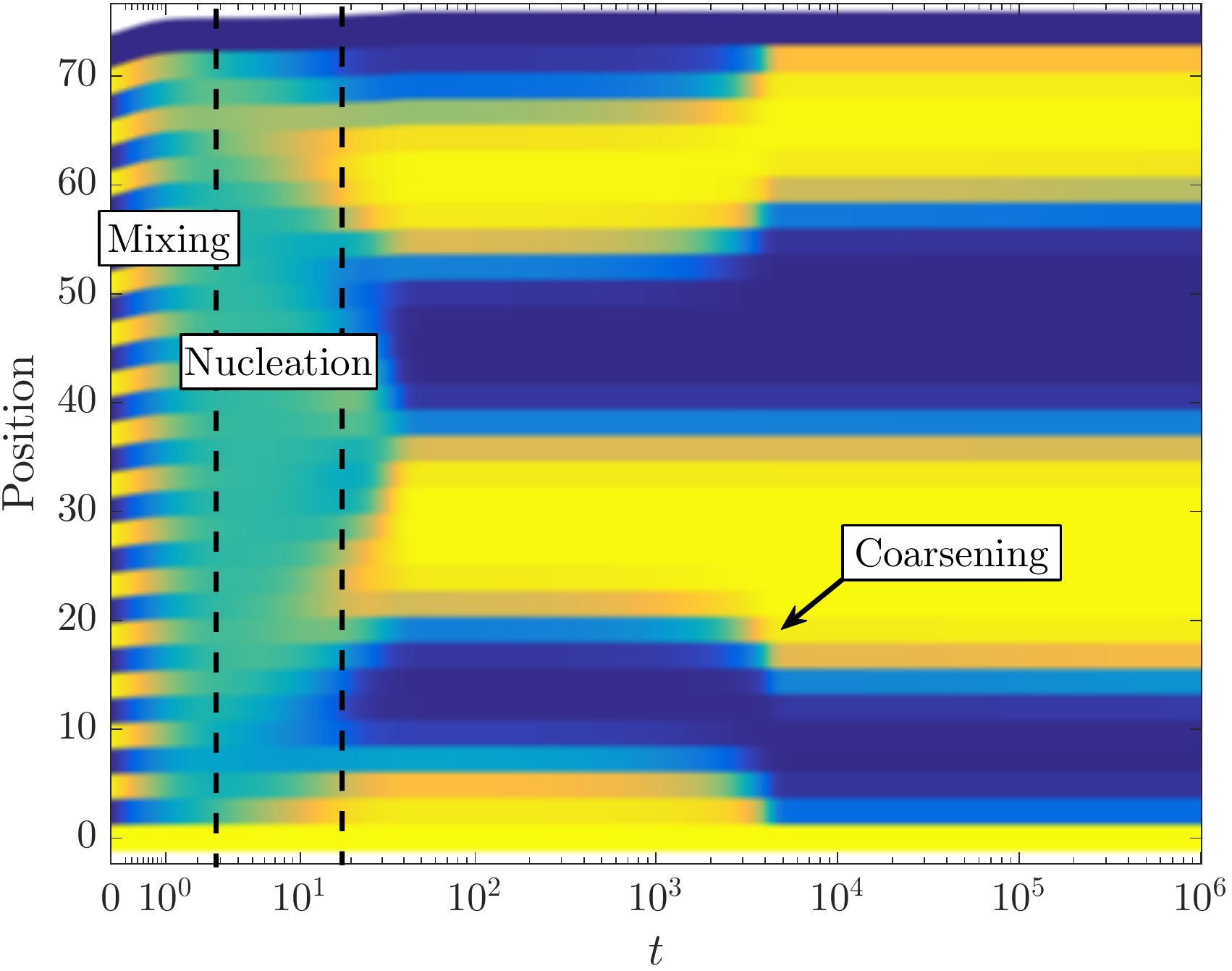}}
  \subfigure[Free Energy - Constant
  Mobility]{\includegraphics[width=6.25cm]{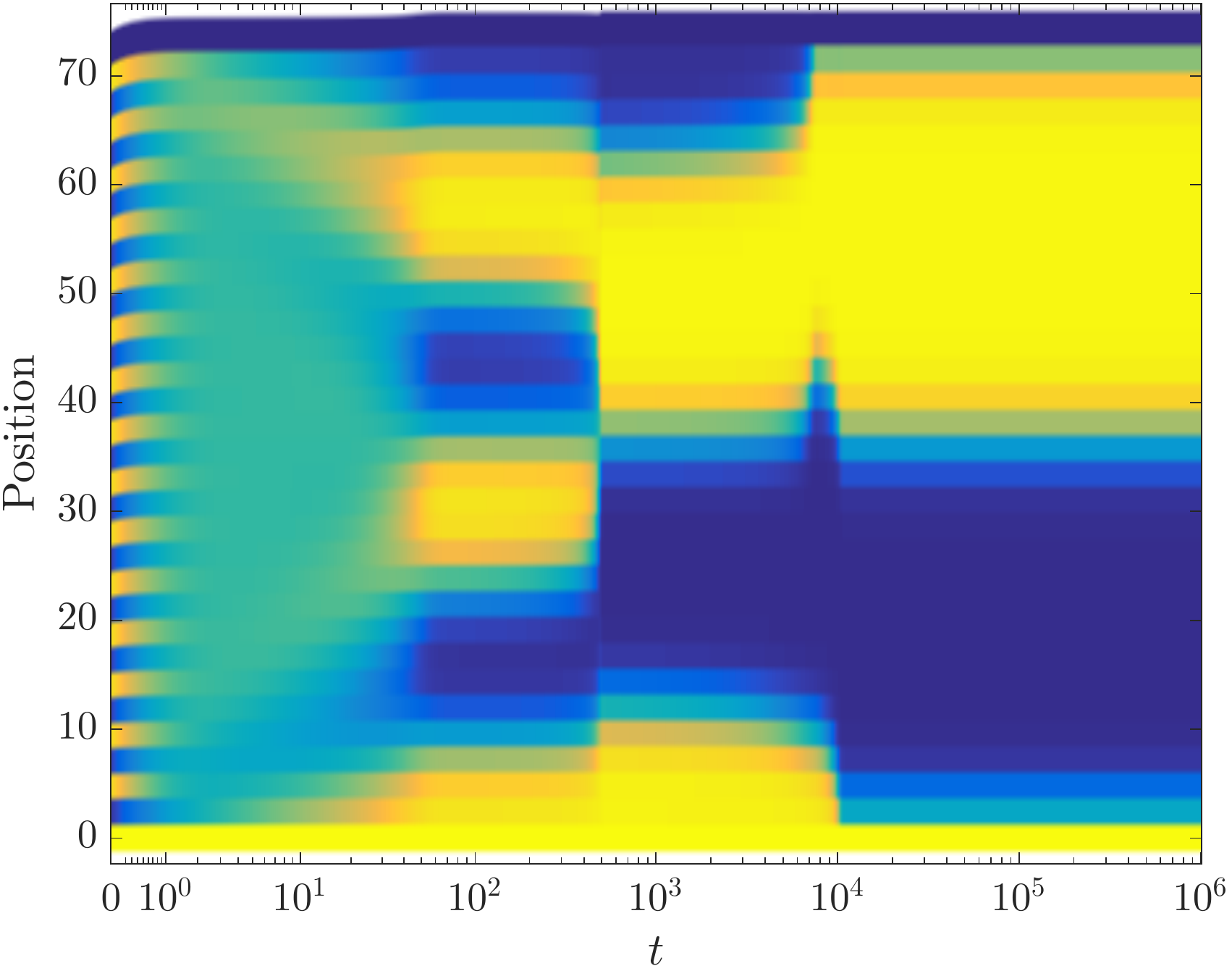}}

  \caption{{A comparison of different dynamical models for a
      system with $N=32$ and $\beta = 160$.  The mean field model with
      point estimates behaved much like (c) and the free energy model
      with rate limited mobility behaved much like (d). }}
  \label{f:N32_beta160_profiles}
\end{figure}

In the case of the small system, $N=32$, the composition and
configuration profiles are shown in
Figure~\ref{f:N32_beta160_profiles}.  The composition dynamics has
three main features:
\begin{enumerate}
\item {\bf Mixing:} Initially alternating positive and negative spins
  become almost constant zero.
\item {\bf Nucleation:} Phases of positive and negative spins form,
  initially at the boundaries, and then in the interior.
\item {\bf Coarsening:} Phases merge, resulting in larger phases.
\end{enumerate}
Each of these changes in the composition occurs in conjunction with
lengthening of the configuration. These features are shown in
Figure~\ref{f:N32_beta160_profiles}.  In, for instance,
Figure~\ref{f:N32_beta160_profiles} (c), we see mixing occurs between
$t=0$ and $t \approx 1$, nucleation follows between $t \approx 1$ and
$t \approx 10$, and then a coarsening event occurs around
$t = 5 \times 10^3$.

\begin{figure}
  \centering \subfigure[Evolution of
  strain]{\includegraphics[width=6.25cm]{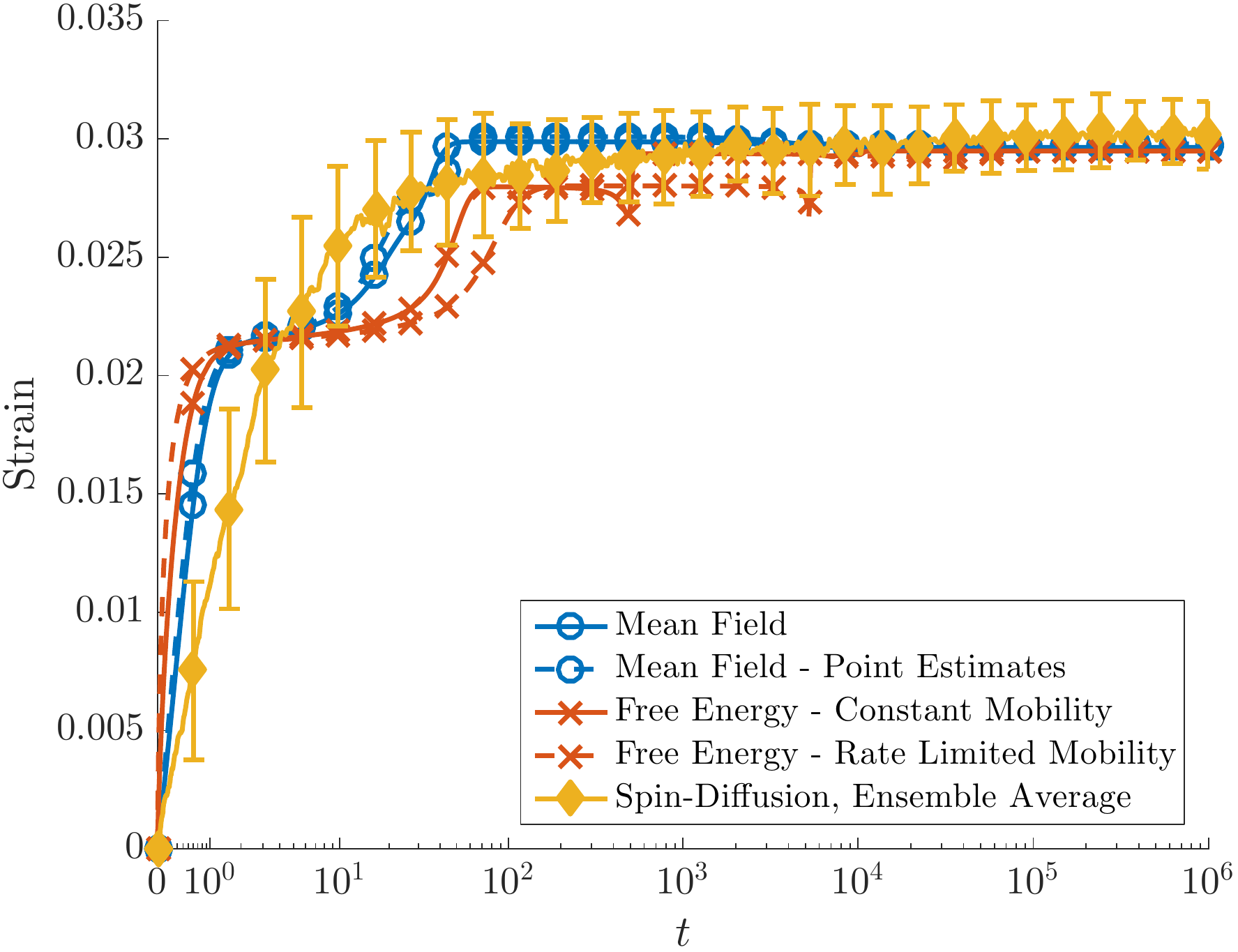}}
  \subfigure[Evolution of the A-B bond
  fraction]{\includegraphics[width=6.25cm]{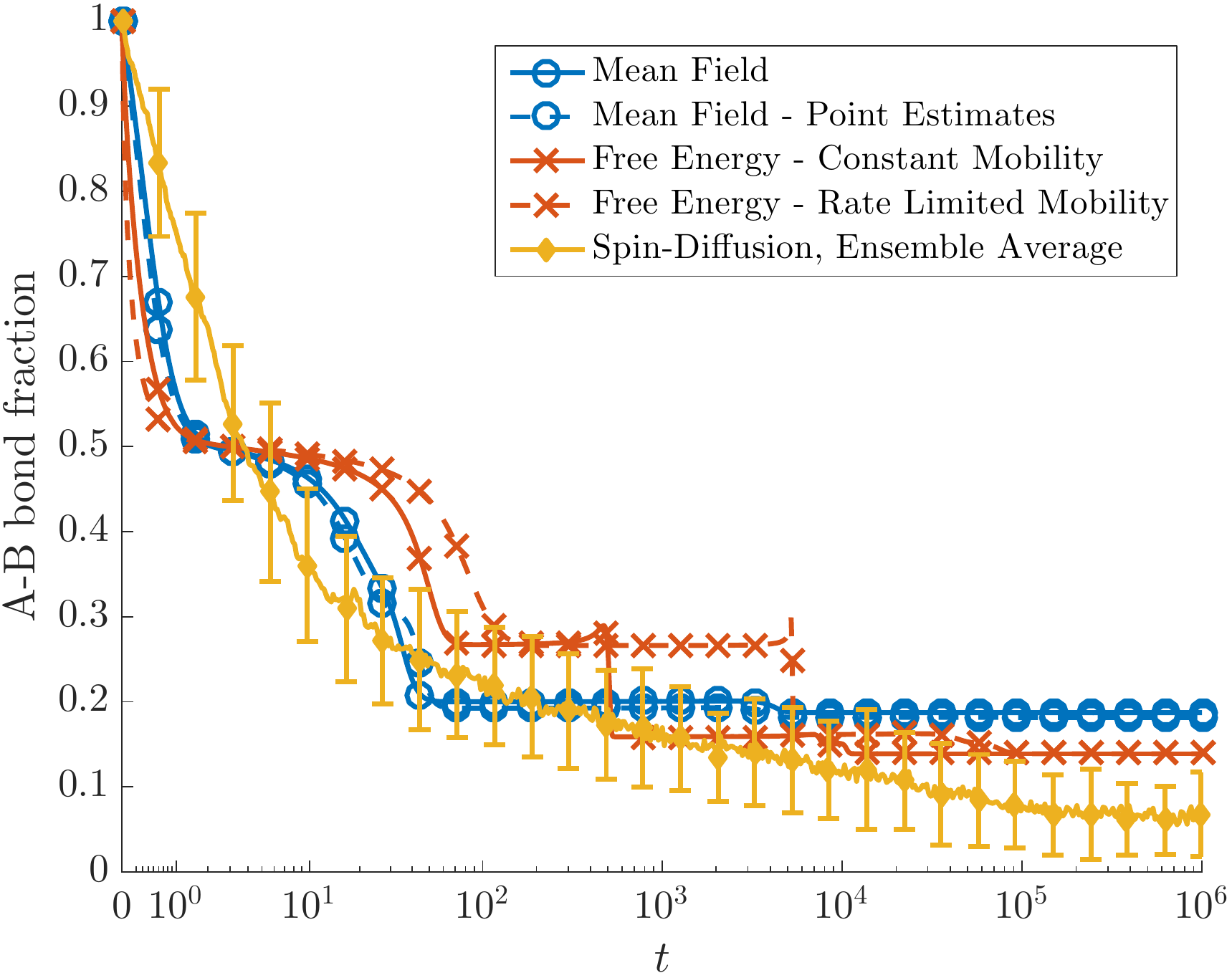}}

  \caption{{Observables for the $N=32$ system at $\beta = 160$.
      The stochastic ensemble average is plotted with $\pm 1$ standard
      deviations.}}
  \label{f:N32beta160observables}
\end{figure}

{In this small system, there are a few coarsening events as the
  system tends towards total segregation of the two species.  Near
  total segregation is visible in the ensemble average,
  Figure~\ref{f:N32_beta160_profiles} (b), while the deterministic
  simulations exhibit incomplete coarsening.  It is clear that
  nucleation completes earlier for the mean field master equation
  model than for free energy gradient dynamics. For $N=32$,
  $\beta=160$, nucleation completes around $t=10$ for the mean field
  models, but closer to $t=10^2$ for the free energy gradient
  dynamics. Another distinction is that the mean field models have
  better defined grains, with sharper interfaces at this temperature.
  The stages of mixing and nucleation are more difficult to
  distinguish in the ensemble average, but we do see a coarsening
  event there around $t=10^4$.

  Differences between the models are also visible when we examine the
  strain and fraction of A-B bonds as functions of time; these appear
  in Figure \ref{f:N32beta160observables}.  Observables for the
  stochastic model evolve without plateauing.  We believe the plateaus
  are a consequence of the mean field approximation and and the use of
  a system with comparatively short range interactions in the model at
  hand.  Despite this discrepancy, a large fraction of each of the
  deterministic trajectories is within one standard deviation of the
  ensemble averages.

  After mixing occurs, the system has a more rapidly increasing strain
  and decreasing number of A-B bonds in the mean field approximations
  than in the free energy gradient dynamics.  The strain fields all
  agree in the long time limit, and, while there is still some
  disagreement in the fraction of A-B bonds, as the system tends to
  full segregation, these too will agree amongst the models.}

\subsubsection{Large Systems}

\begin{figure}
  \centering \subfigure[Single
  Realization]{\includegraphics[width=6.25cm]{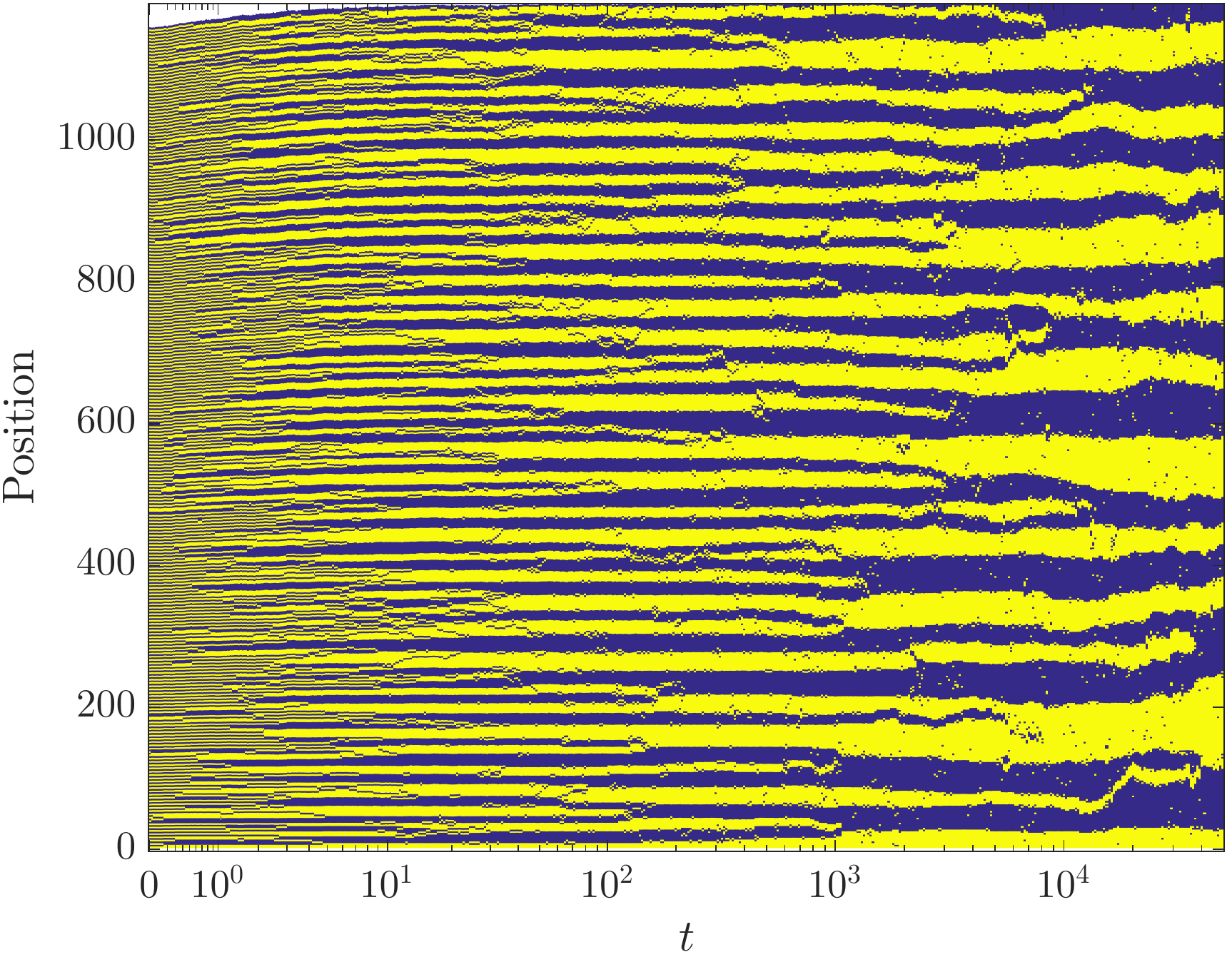}}
  \subfigure[Ensemble
  Average]{\includegraphics[width=6.25cm]{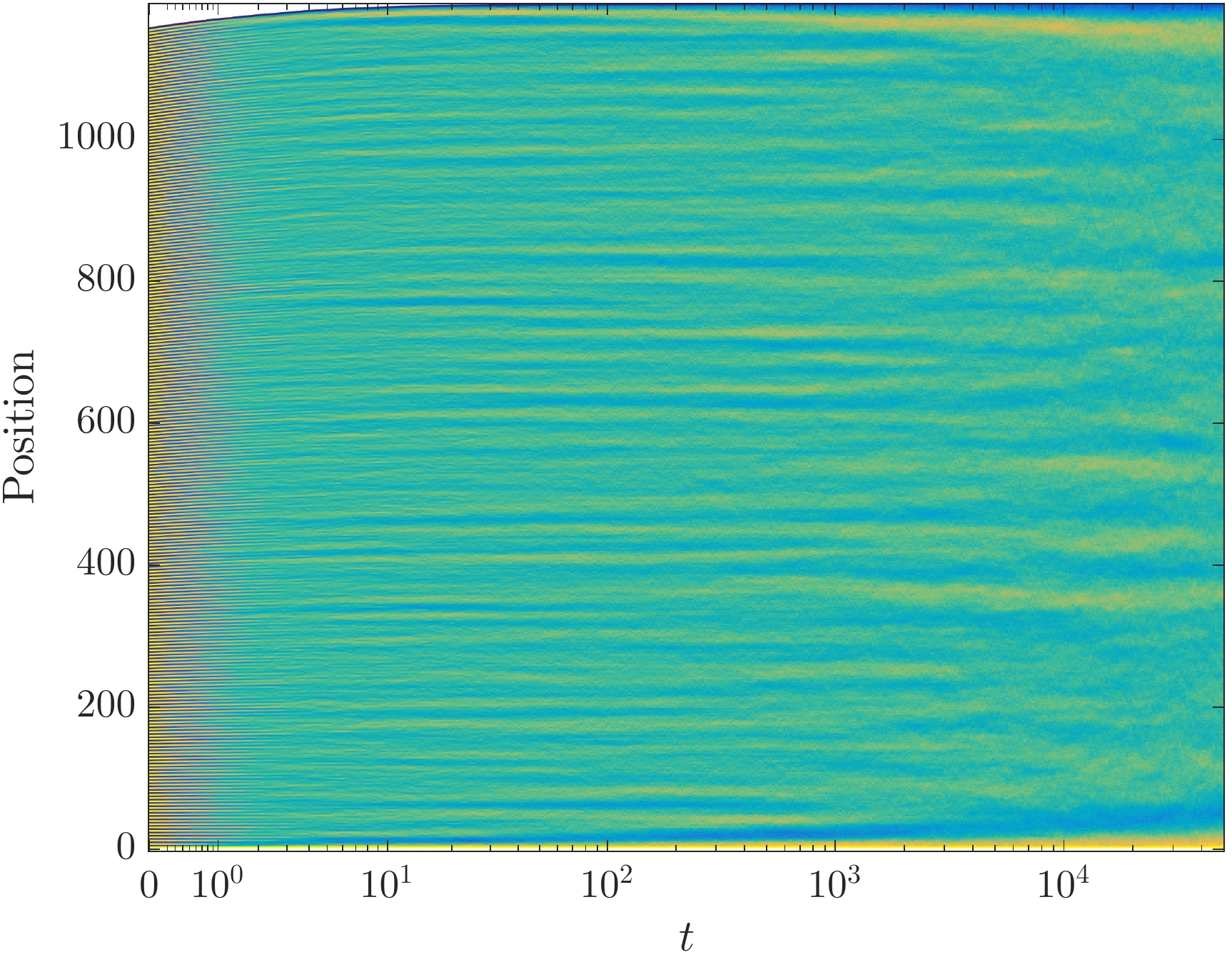}}

  \subfigure[Mean
  Field]{\includegraphics[width=6.25cm]{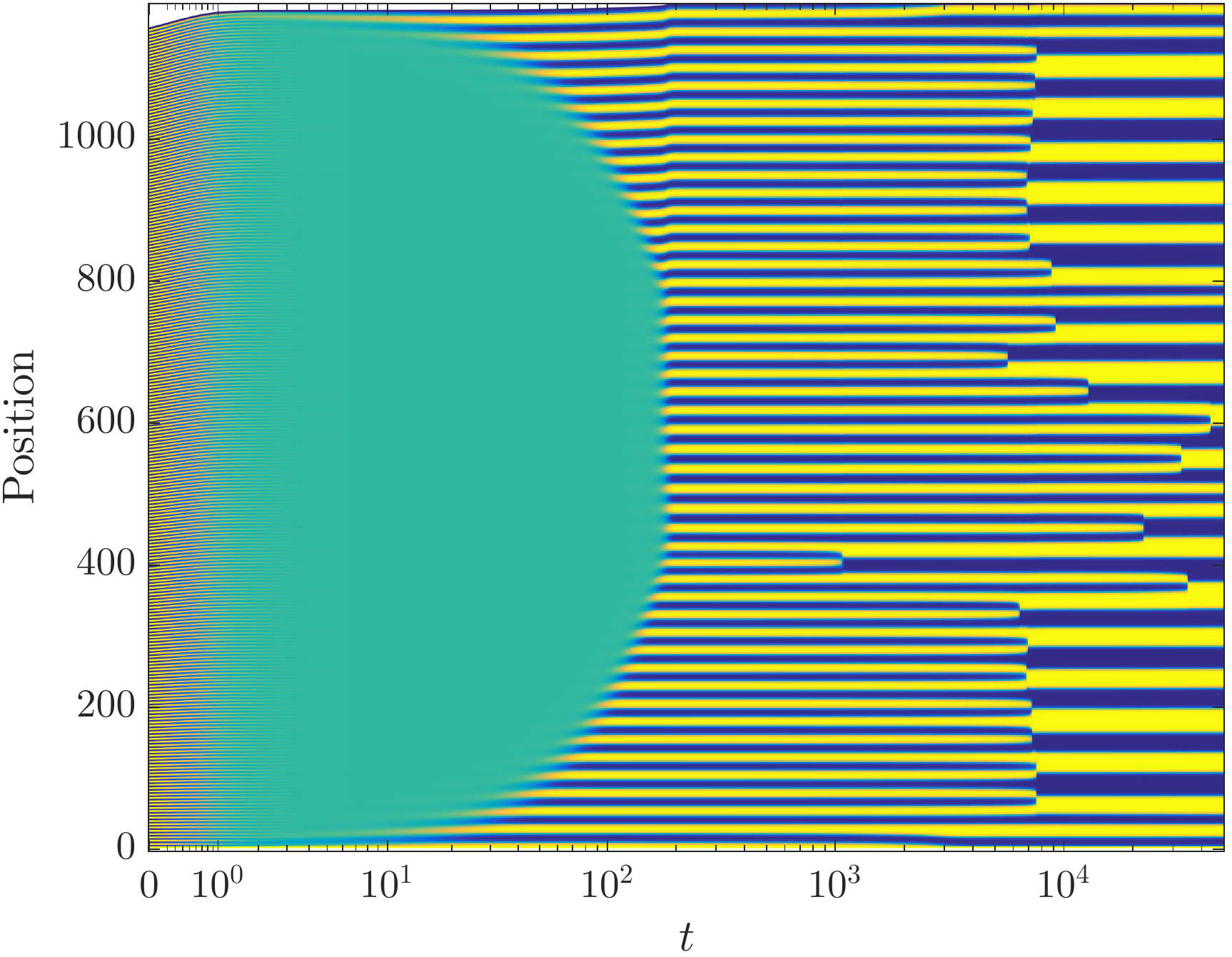}}
  \subfigure[Free Energy - Constant
  Mobility]{\includegraphics[width=6.25cm]{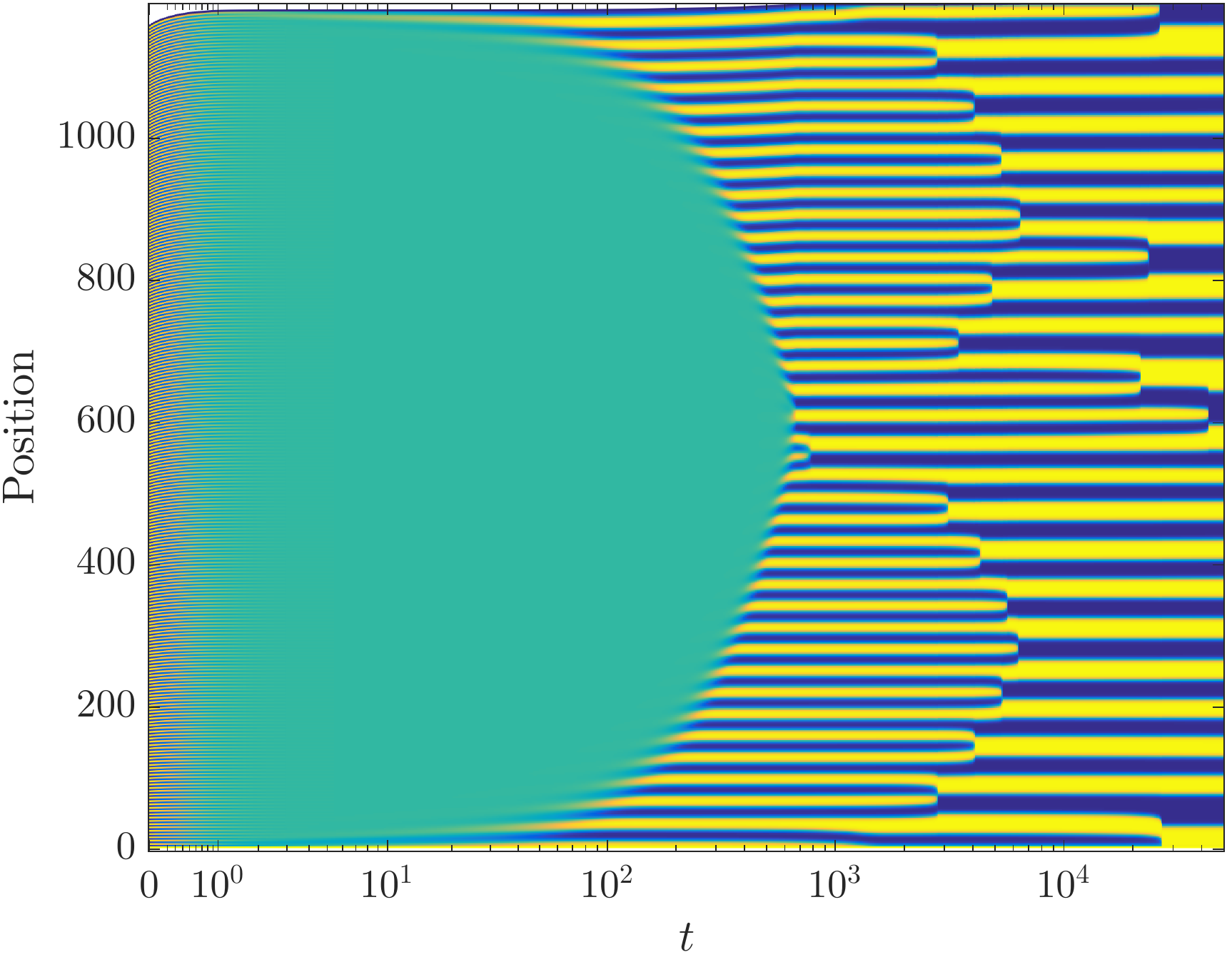}}

  \caption{{A comparison of different dynamics for a system with
      $N=500$ and $\beta = 160$.  Compare with the $N=32$ case, Figure
      \ref{f:N32_beta160_profiles}. }  }

  \label{f:N500_beta160_profiles}

\end{figure}

\begin{figure}
  \centering
  \subfigure[]{\includegraphics[width=6.25cm]{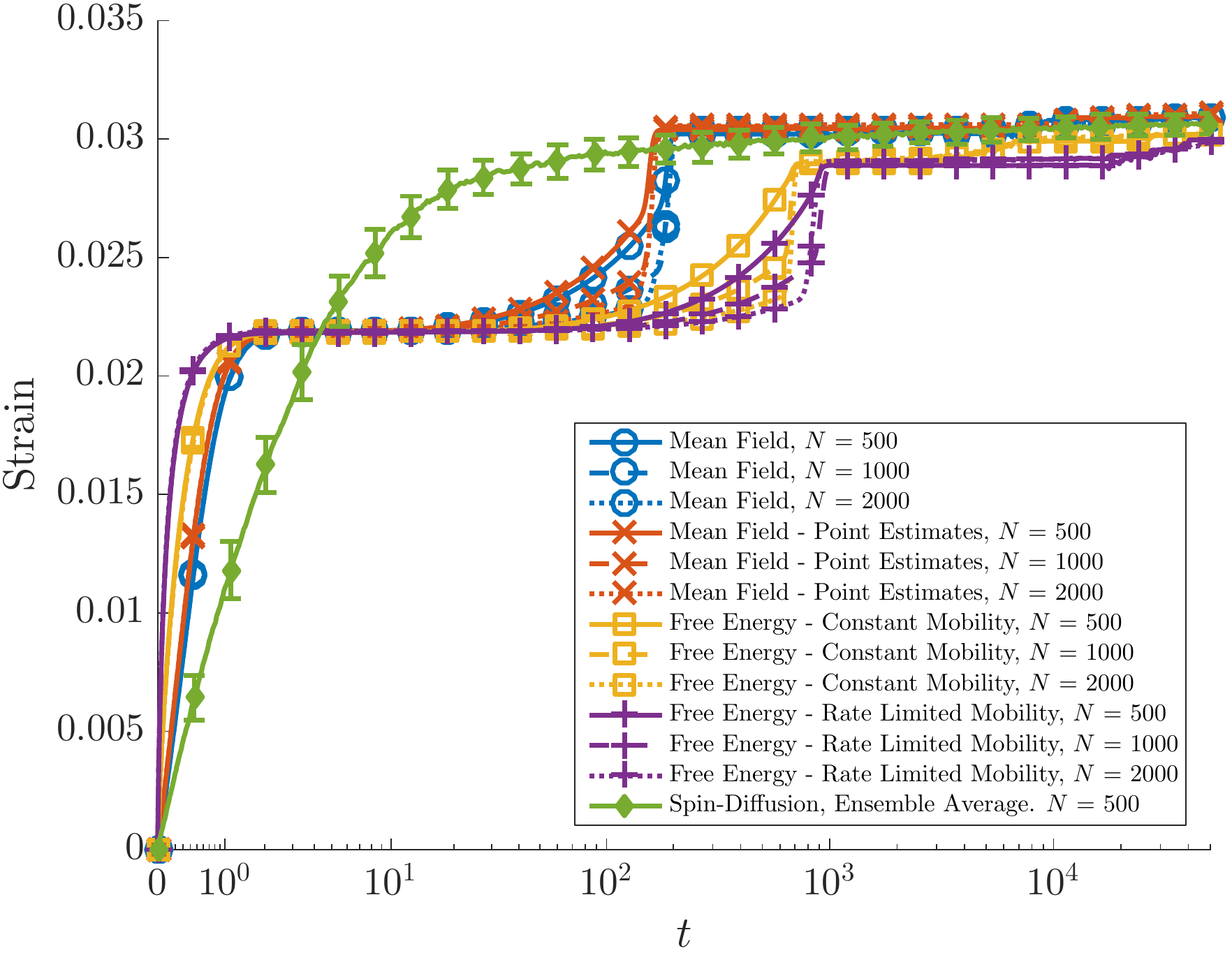}}
  \subfigure[]{\includegraphics[width=6.25cm]{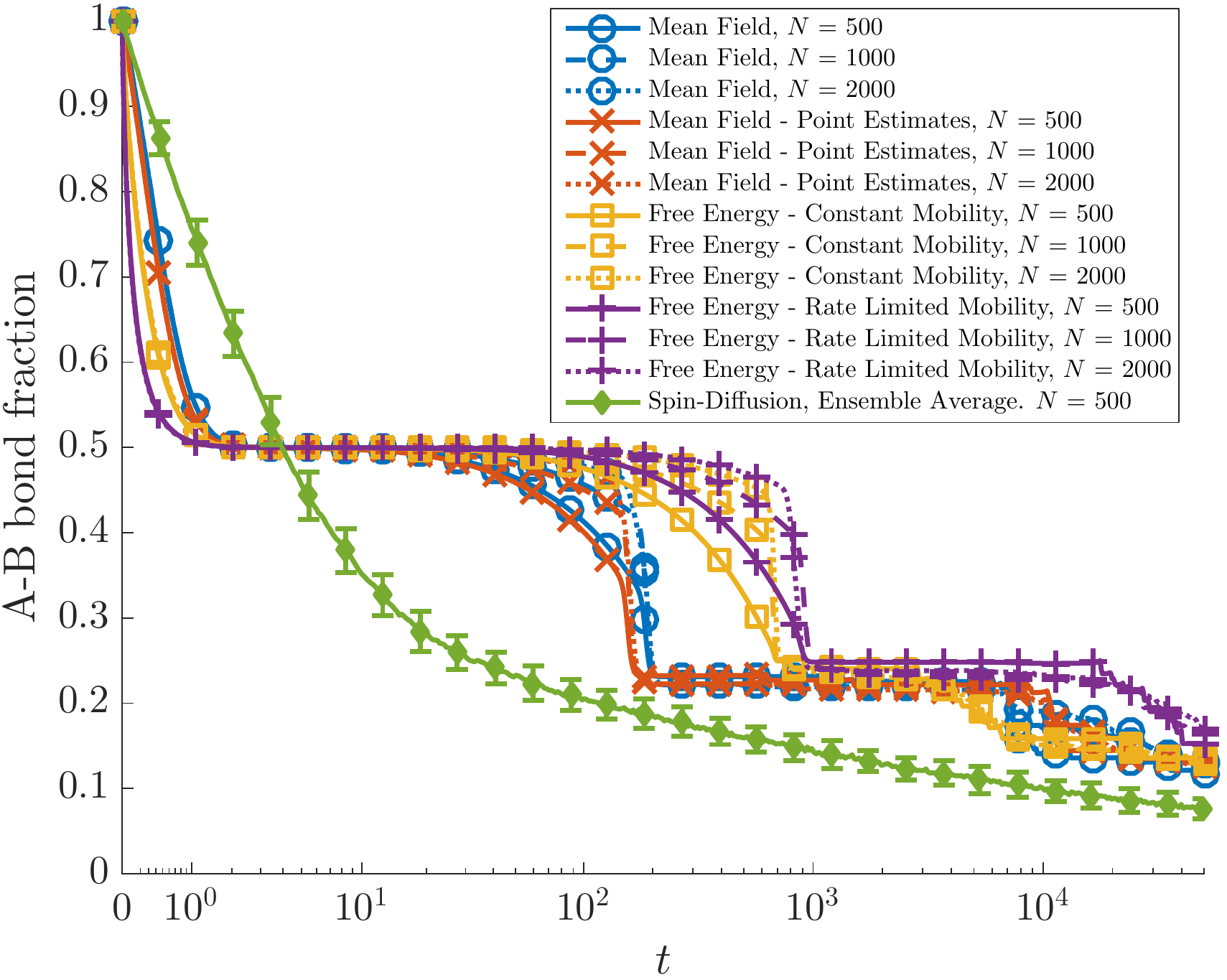}}

  \caption{Observables for deterministic models at $N=500,1000,2000$
    systems at $\beta = 160$, along with the ensemble average at
    $N=500$ with $\pm 1$ standard deviation.  Compare with the $N=32$
    case in Figure \ref{f:N32beta160observables}.}

  \label{f:Nlarge_beta160_observables}

\end{figure}

{We also investigated the DMD model in larger systems. The time horizon of the simulated dynamics was limited to
  $t=5\times 10^4$ due to the computational complexity of the
  stochastic model.  While such time horizon is too short to see total
  segregation, we are still able to observe nucleation, mixing, and
  multiple coarsening events.

  Looking at the autocorrelation in Figure~\ref{f:N32beta160scaling}
  (b), we see coarsening of the system in all cases.  Turning to
  Figures~\ref{f:N500_beta160_profiles} (c) and (d), we see that, as
  in the smaller system, nucleation completes earlier for the mean
  field model, around $t=10^2$, than for free energy gradient
  dynamics, around $t=10^3$.  A large number of coarsening events
  occur between $t=10^3$ and $t=10^4$ for the free energy gradient
  model, about an order of magnitude earlier than in the mean field
  model.

  There are much less well defined phases in the ensemble average, as
  seen in Figure~\ref{f:N500_beta160_profiles} (b), though there is
  still coarsening.  In particular, the interfaces between the regions
  of segregated species are less well defined resulting in a
  ``slush.''  This can partly be explained by examining Figure
  \ref{f:N500_beta160_profiles} (a), where we see there is uncertainty
  as to where the exact locations of the interfaces are, resulting in
  an averaging out.

  Turning to observables in Figure \ref{f:Nlarge_beta160_observables},
  we see disagreement between the ensemble average and the
  deterministic models.  The smaller $N=32$ system, in comparison,
  showed better agreement between modes; see Figure
  \ref{f:N32beta160observables}.

  Not only are the plateaus absent, but the deterministic dynamics are
  far from the distribution of the stochastic ensemble.  We posit that
  this is a consequence of the mean field approximation.

  There is consistency across the deterministic models of different
  sizes $N=500,1000,2000$ in the long time limit of the observables.
  For strain, there is even agreement with the ensemble average of
  stochastic trajectories.  The long time behavior of the stochastic
  model is less consistent in the A-B bond fraction observable, but we
  conjecture we would observe improved agreement as the segregation
  continues and eventually completes.}

\section{Discussion}
\label{s:discussion}

We have formulated and related an underlying stochastic process, the
coupled spin-diffusion model, to DMD, and shown that there is
qualitative agreement in simulations.  Our simulations also show that,
depending on the modeling parameters, there are differences in the
dynamics.  Our contention is that, given that such distinctions occur,
practitioners should follow the strategy developed here, and put their
key modeling assumptions in the reaction rates and the mean field
approximations.

\begin{figure}
  \centering
  \includegraphics[width=6.25cm]{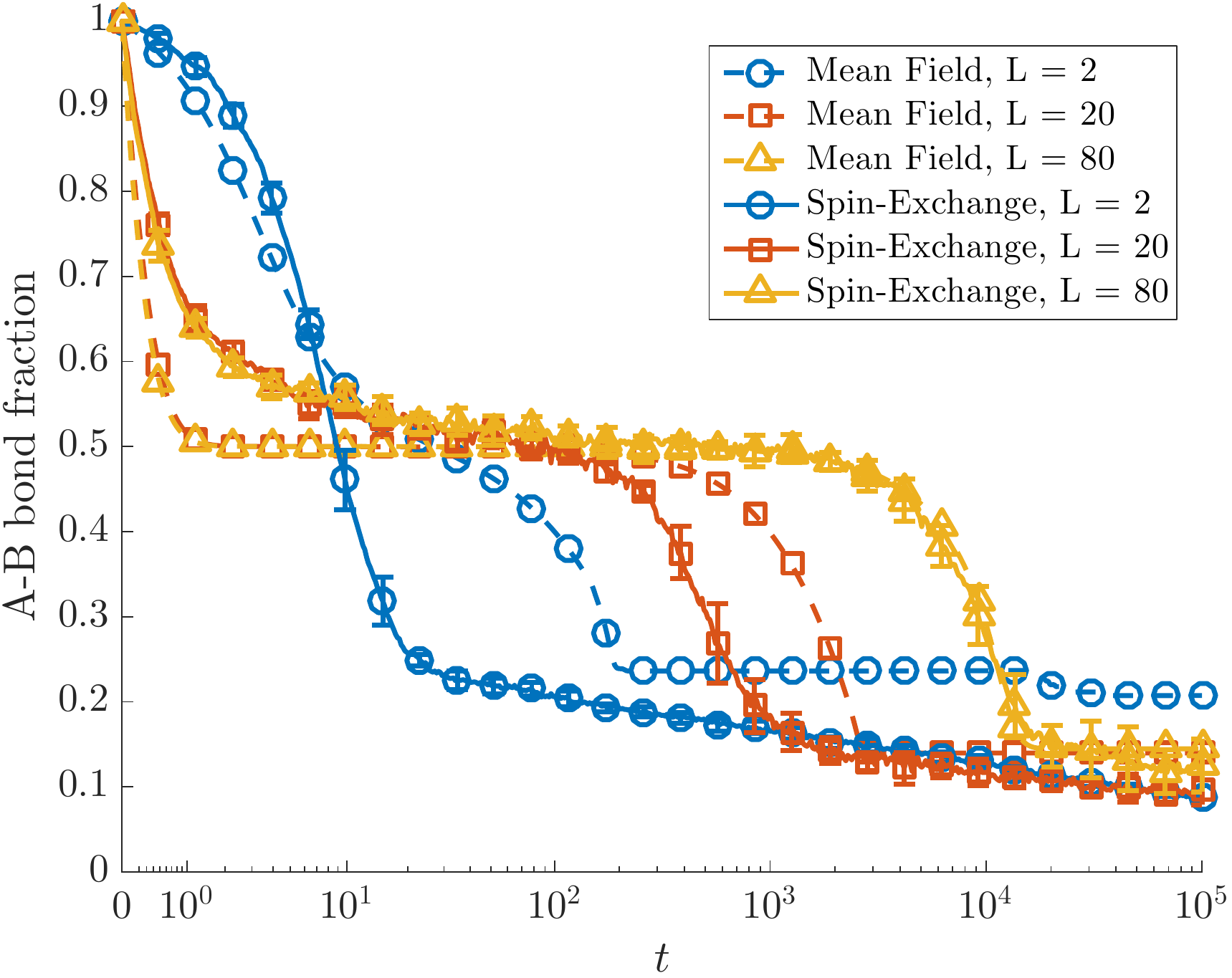}

  \caption{Evolution of an on-lattice Ising-type model with the energy
    \eqref{e:ising} at $\beta = 2$ with $N =1000$.  Compare with
    Figure~\ref{f:Nlarge_beta160_observables}.}

  \label{f:ising}

\end{figure}

{There is clear disagreement between the ensemble average of the
  stochastic model and the deterministic model, in that the ensemble
  averages have less sharply defined interfaces between phases, as in
  Figures~\ref{f:N500_beta160_profiles}, and the observables do not
  plateau as in the deterministic model, as in Figure
  \ref{f:Nlarge_beta160_observables}.  We believe this is the result
  of studying a system in one dimension with comparatively short range
  interactions.  If we look at a system with an ABABAB...
  composition, the values of the $J_{ij}$ in \eqref{e:modelV2} are as
  in Table~\ref{t:Jij}.  These are obtained from the equilibrium
  displacements obtained after quenching.  The values indicate that
  while the system certainly has the first and second nearest neighbor
  interactions, the third nearest neighbors are comparatively weak,
  and beyond the third, there is negligible interaction.

  \begin{table}
    \caption{{Values of the interaction coefficients in
        \eqref{e:modelV2} for an ABABAB... composition with
        displacements computed at zero temperature.}}
    \label{t:Jij}
    \centering
    \begin{tabular}{l r r}
      \hline\hline
      Nearest Neighbor & $r_{ij}$ & $J_{ij}$\\
      \hline
      First & 2.28 &0.0026\\
      Second &4.56 &0.0036\\
      Third &6.84& 0.0009\\
      Fourth & 9.12& 0.0001\\
      \hline
    \end{tabular}
  \end{table}

  In comparison, the mean field approximation of the on lattice Ising
  model is not expected to be accurate in one dimension, with such
  short range interactions, \cite{Chandler:1987aa,
    Baxter:1989ug,Presutti:2009ha,Ellis:2012wd}.  Indeed,
  Figure~\ref{f:ising} plots the time evolution for the on-lattice
  analog of the test problem presented in Section~\ref{s:numerics}.
  Here, the energy is,
  \begin{equation}
    \label{e:ising}
    V(\bsigma) = -\frac{1}{2L}\sum_{i=1}^N\sum_{0<|i-j|\leq L} \sigma_i \sigma_j,    \quad \sigma_j =
    \begin{cases}
      1, & \text{for $j \leq 1$,}\\
      -1, & \text{for $j \geq N$}.
    \end{cases}
  \end{equation}
  We see similar discrepancies to what was observed in Figure
  \ref{f:Nlarge_beta160_observables}.  We expect such discrepancies
  will be mitigated in higher dimensional problems where the number of
  interacting sites increases and longer range potentials are applied.

  This serves as a warning for the practitioner who seeks to use DMD
  type models. If the system in question only has short range
  interactions, such discrepancies will occur.  It is an important
  problem to formulate and explore 2D and 3D test problems for these
  models, and to examine the limitations of the mean field
  approximation.  We do expect agreement in the observables in the
  long time limit.  Thus, even if the detailed temporal evolution,
  e.g., the transient events, is incorrect, the deterministic model
  is more easily integrated to equilibrium than the corresponding
  stochastic model.  }

We summarize the key assumptions and approximations that are at the
core of the studied model.
\begin{description}
\item[Spin-Diffusion Model] The underlying modeling assumption is that
  the system is governed by the coupled system of \eqref{e:diffusion1}
  and \eqref{e:rates1}.

\item[Time Scale Separation] The next assumption is that there is
  adequate scale separation between the relaxation time and the time
  scale for barrier crossings.  This is captured by Assumption
  \ref{asm:qsdtime}, which is the basis for Approximation
  \ref{apx:leadingqsd}, the rate averaged jump process.

\item[Low Temperature Approximation] To make the computation of the
  effective reaction rates tractable, the QSD is approximated by the
  restricted Gibbs distribution (Approximation \ref{apx:lowtempqsd}),
  which in turn may be approximated by a Gaussian distribution using
  VG.  This is sensible in the low temperature limit, and was captured
  by \eqref{e:comp1} in our comparison of the models.

\item[State Space Approximation] The first major approximation is in
  Approximation \ref{apx:statespace}, which restricts our attention to
  a single new basin each time a spin-exchange occurs.

\item[Mean Field Approximation] The final approximation, giving rise
  to equations like \eqref{e:qsd_mfmaster1}, is to approximate
  expectations of nonlinear functions by the nonlinear functions of
  the expectations.
\end{description}
The time scale separation and low temperature approximations are quite
sensible, are consistent with data, and they can be made rigorous.  At
present, the state space and mean field approximations are somewhat
less constrained.  The mean field approximation is quite common, and
can even be justified in certain limits. The state space approximation
merits further investigation.

One interpretation of our model is as a mean field approximation to a
master equation for an off-lattice, finite temperature, restricted kMC
method.  It is clearly off-lattice as the atomic sites evolve in time,
and it certainly has finite temperature effects throughout. We view it
as a kMC model because the system is driven by the evolution of the
composition, {\it i.e.}, the assignment of the A and B species to
different sites.

Indeed, one might arrive at our model by beginning with, for instance,
the off-lattice kMC model of \cite{Boateng:2014il}. Instead of viewing
the positions of the atoms as corresponding to their quenched
configurations, one could allow them to sample the local mode of the
energy landscape, as is done in accelerated MD
approaches~\cite{perez2009accelerated,Voter:2002p12678}.  There, the
system reaches a local equilibrium, sampling a QSD, and then one of
several approaches is used to accelerate the observation of a first
exit.  As noted, another difference between DMD type methods and the
method of \cite{Boateng:2014il} is that the latter produces single
realizations of the trajectories, requiring ensemble averaging.

{This difference is also found when comparing with accelerated MD
  methods.  Accelerated MD can efficiently produce single
  trajectories, but will require ensemble averaging to predict the
  evolution of observables.  Another difference is that accelerated MD
  methods allow for all types of transitions amongst basins, while DMD
  privileges species exchange, which correspond to correlated events.
  This restricts the underlying kMC model of DMD.  Other off-lattice
  kMC methods, such as those in \cite{Xu:2008jk,Xu:2009fy}, are
  constructed from transitions between adjacent basins of $V(\bx)$
  connected by a single saddle point.  The exchanges of species would
  not be captured by single kMC event in such a model, requiring
  multiple transitions.}

{There are, of course, many challenges ahead.  As noted, an
  important task is to make direct numerical comparisons between
  ensemble averages of the spin-diffusion model and the deterministic
  model in other test problems.  In particular, it would be essential
  to look at problems in two and three dimensions with longer range
  potentials, where the mean field approximations is expected to be
  more consistent. This would allow for an assessment of the
  aforementioned approximations, and regions of validity could be
  determined. Alternatively, there may be ways to correct the
  mean field approximation.}

\appendix

\section{Details of Potentials}
\label{a:modelV}

First, we give the details for the coefficients $J_{ij}$, $h_i$, and
$f$. We have
\begin{align*}
  J_{ij}(\bx) &= \frac{1}{4} \left[ 2 \phi_{+-}(|x_{ij}|) - \phi_{++}(|x_{ij}|) - \phi_{--}(|x_{ij}|) \right],\\
  h_i(\bx) &= \frac{1}{4} \sum_{j \neq i} \left[ \phi_{++}(|x_{ij}|) - \phi_{--}(|x_{ij}|) \right] + \frac{1}{2} \left[ u_+(|x_i|) - u_-(|x_i|) \right], \text{ and}\\
  f(\bx) &= \frac{1}{4} \sum_{i<j} \left[ \phi_{++}(|x_{ij}|) + 2 \phi_{+-}(|x_{ij}|) + \phi_{--}(|x_{ij}|) \right] + \frac{1}{2} \sum_i \left[ u_+(|x_i|) + u_-(|x_i|) \right].
\end{align*}

For the pair interactions in our test problem, we use a soft-core
$4$-$2$ Lennard-Jones type potential:
\begin{equation}
  \label{e:ljsmooth}
  \phi(r; A, r_{\rm eq}, \lambda) =
  4 \lambda^2 A \left\{ \left[ \tfrac{1}{2} (1 - \lambda)^2 + 2 \left( \tfrac{r}{r_{\rm eq}}\right)^2 \right]^{-2} -  \left[ \tfrac{1}{2} (1 - \lambda)^2 + 2 \left( \tfrac{r}{r_{\rm eq}}\right)^2 \right]^{-1} \right\}.
\end{equation}
We introduce a cutoff of the above pair potential, to obtain a pair
potential that is $C^1$ with finite support in $[0, r_{\rm c}]$:
\begin{equation}
  \label{e:cutoff}
  \phi_{\mathrm{cut}}(r; p, A, r_{\rm  eq}, \lambda, r_{\rm c}) =
  \begin{cases}
    \phi(r) - \phi(r_{\rm c}) - \phi'(r_{\rm c}) (r - r_{\rm c}), & r < r_{\rm c},\\
    0, & r \geq r_{\rm c}.
  \end{cases}
\end{equation}
In addition to the pair potential, there is a quartic confining
potential to prevent evaporation occurring as a rare event:
\begin{equation}
  \label{e:confine}
  \phi_{\mathrm{confine}}(r; r_{\mathrm{confine}}) = \left[  \tfrac{1}{4} \left( \tfrac{r}{r_{\mathrm{confine}}} \right)^4 - \tfrac{r}{r_{\mathrm{confine}}} + \tfrac{3}{4} \right] 1_{(r_{\mathrm{confine}}, \infty)}(r).
\end{equation}

The potential parameters are $\lambda = 0.99$,
$A_{++} = A_{--} = 0.2$, $A_{+-} = 0.18$,
$r_{\eq,++} = r_{\eq,--} = 2.6$, $r_{\eq,+-} = 2.55$. The cutoff
radius is $r_c = 10.5$, and the confining potential has parameter
{$r_{\mathrm{confine}} = 2 \cdot r_{\eq,+-}= 5.1$}.

\section{Details of Numerical Methods {for Deterministic
    Models}} \label{a:numerics}

In this Appendix, we provide further details about the numerical
computation of the free energy, as well as other computational
parameters for the simulations.  Though the presentation is in terms
of $\bX$, $\bk$, and $\bs$, the simulations were performed in terms of
$\bX$, $\bm{\alpha} = \beta\bk/2$ and $\bc = (\bs + 1)/2$.

\subsection{Calculation of free energy and its derivatives}

Many of the expressions require computing expectations over a
variational Gaussian distribution with parameters $\bX$ and
$\bm{\alpha}$.  Since
$\E^{\tilde{\nu}} \left[ V(\bx, \bsigma) \right] = \E^{\tilde{\mu}}
\left[ V(\bx, \bs) \right]$, for pair potential averages, like
$\E^{\tilde{\mu}} \left[ \phi(|x_{ij}|) \right]$, we can use the
coordinate transformation from \cite{Li:2011gn} which makes the
integral one dimensional.  We also integrate over a truncated region,
$(X_i-R, X_i+R)$, in one dimension, with $R$ chosen sufficiently large
that the domain truncation error is within a specified tolerance.
This is done with {\it a priori} estimates involving the maximum value
of the pair potential, and an assumed lower bound $\alpha_{\min}$ on
the harmonic constants $\bm{\alpha}$.

The confining potential applies only to nearest-neighbor pairs, so we
do not need a cutoff and therefore do not need to truncate the domain
of integration for computation of its Gaussian average.

\subsection{Computational Parameters}

There are several computational parameters for the calculation of the
free energy and its gradients. First, there is $R$, the radius of
truncation described above. In our simulations, $R \approx 3.7$, which
is a function of the {\it a priori} lower bound $\alpha_{\min} = 5$,
the upper bound
$M = \max \{ \phi_{++} (0), \phi_{+-} (0), \phi_{--} (0) \}$, and the
truncation tolerance of $10^{-6}$.  The absolute and relative
tolerances for the adaptive quadrature have values $10^{-12}$ and
$10^{-6}$ respectively.

There are also computational parameters for the minimization of
$\mathcal{F} (\bc, \bX, \bm{\alpha})$ by gradient descent. The
absolute and relative tolerances for the ODE solver are $10^{-6}$ and
$10^{-3}$, respectively. The time step is chosen adaptively to meet
these error tolerances. The gradient descent is stopped when
$\max( \|\nabla_{\bX} \mathcal{F} \|_{\infty}, \|\nabla_{\bm{\alpha}}
\mathcal{F} \|_{\infty})$ falls below the tolerance $10^{-7}$.

Finally, we have several computational parameters for evolution of the
$\bc$, $\bX$, and $\bm{\alpha}$ by the transformed versions
of~\eqref{e:ds_dt} -- \eqref{e:dk_dt}. The absolute and relative
tolerances for the ODE solver are $10^{-6}$ and $10^{-3}$,
respectively. The evolution is halted when the infinity-norm of the
gradient increases above the tolerance $10^{-6}$. The evolution of
these equations continues once $\bX$ and $\bm{\alpha}$ have been
updated by gradient descent minimization. {Finally, the
  prefactors on the evolution of $\bX$ and $\bm{\alpha}$ are both
  $10^{10}$, which is equivalent to $\eta_1 = 10^{10}$ and
  $\eta_2 = (2/\beta)^2 10^{10}$ in~\eqref{e:dX_dt}
  and~\eqref{e:dk_dt}.}

\section{{Details of Numerical Methods for Spin-Diffusion
    Models}} \label{a:numerics_spinDiff}

{ In this Appendix, we provide details of the numerical
  computation of spin-diffusion trajectories. We simulate the leading
  order process in Approximation~\ref{apx:leadingqsd}, with exchange
  rates~\eqref{e:tanh1}, and the State Space
  Approximation~\ref{apx:statespace}.  The QSD averages of exchange
  rates and observables are estimated with time averages.  These are
  computed by integrating the diffusion with preconditioned Metropolis
  Adjusted Langevin (MALA), \cite{Roberts:1996tm}.  While not strictly
  applicable to QSD sampling, in our low temperature regime, this
  efficiently produced consistent results. We obtained one hundred
  trajectories of the spin-diffusion process, from which we can
  estimate ensemble averages and confidence intervals.  }

\subsection{{Basin Sampling via Preconditioned MALA}}

{For a given spin state, the basin is determined by finding the
  local energy minimizer from the last known minimizer.  In this way,
  we capture Approximation~\ref{apx:statespace}. Energy minimization
  is performed using the {\sc Matlab} function \texttt{fminunc}.
  Given the minimizer $\bx_{\min}$ of the new spin state and basin, we
  use it to initialize preconditioned MALA with the inverse of the
  Hessian $H(\bx_{\min})$ used as the preconditioner.  The time step
  is chosen to maintain an acceptance rate between 65\% and 85\%.
  Estimates of the exchange rates and observables are obtained by
  averaging over the samples produced by preconditioned MALA.}

\subsection{{Computational Parameters}}

{ The tolerances for minimization of $V (\bx)$ by the trust
  region method are $10^{-6}$ for the norm of the gradient, $10^{-6}$
  for the relative change in the energy, and $10^{-8}$ for the norm of
  the change in $\bx$.

  The time step for MALA is $\Delta t = 0.2$ for $N=32$ and
  $\Delta t = 0.1$ for $N=500$. The interval of time averaging is
  $t_{\max}=1000$ for $N=32$ and $t_{\max}=500$ for $N=500$. }

\bibliographystyle{siam} \bibliography{spin}

\end{document}

%% file: macrosDMD.tex
\newcommand{\paren}[1]{\left(#1\right)}
\newcommand{\bracket}[1]{\left[#1\right]}
\newcommand{\set}[1]{\left\{#1\right\}}

\newcommand{\R}{\mathbb{R}}        

\newcommand{\eps}{\epsilon}
\newcommand{\E}{\mathbb{E}}        
\renewcommand{\P}{\mathbb{P}}      

\newcommand{\bx}{\mathbf{x}}                   
\newcommand{\bz}{\mathbf{z}}                   
\newcommand{\bsigma}{\boldsymbol{\sigma}}      
\newcommand{\blambda}{\boldsymbol{\lambda}}    
\newcommand{\bX}{\mathbf{X}}                   
\newcommand{\by}{\mathbf{y}}
\newcommand{\bc}{\mathbf{c}}
\newcommand{\bs}{\mathbf{s}}

\newcommand{\bk}{\mathbf{k}}
\newcommand{\bW}{\mathbf{W}}
\newcommand{\Qm}{Q_{\rm m}}

\newcommand{\bsig}{\boldsymbol{\sigma}}

\newcommand{\calN}{\mathcal{N}}

\DeclareMathOperator{\bigo}{O}

\DeclareMathOperator{\eq}{eq}

\DeclareMathOperator{\argmin}{argmin}

\DeclareMathOperator{\arctanh}{arctanh}

\newtheorem{approximation}{Approximation}
\newtheorem{assumption}{Assumption}